\documentclass[journal]{new-aiaa}

\usepackage[utf8]{inputenc}
\usepackage{textcomp}
\usepackage{graphicx}
\usepackage{amsmath}
\usepackage{color}
\usepackage[version=4]{mhchem}
\usepackage{siunitx}
\usepackage{longtable,tabularx}
\usepackage{gensymb}
\usepackage{float}

\usepackage{graphicx}
\usepackage{subcaption}
\newlength{\subfigwidth}
\newlength{\subfigwidthb}
\usepackage{threeparttable}
\usepackage{capt-of}
\usepackage{booktabs}
\usepackage{varwidth}

\newcommand{\modif}{\textcolor{black}}

\title{Nanosatellite Design Considerations for a Mission to Explore the Propellant Sloshing Problem \footnote[4]{A version of this paper has been presented at the 2023 AIAA SciTech Forum, National Harbor, MD \& Online, January 23, 2023-January 27, 2023, Session: SATS-07, Small Satellite Technologies, Presentation Title: SPICEsat: A Nanosatellite Mission to Explore the Propellant Sloshing Problem, Control ID: 3772694, Paper: AIAA-2023-1878.}
}

\author{Michael Fogel \footnote{Corresponding Author, Graduate Researcher, Department of Mechanical and Aerospace Engineering, Rutgers University, 98 Brett Road, Piscataway, NJ 08854, USA. AIAA Member, michael.fogel@rutgers.edu}}
\affil{Rutgers University, Piscataway, NJ, 08854, USA}

\author{Snigdha Sushil Mishra \footnote{Graduate Student, Department of Computer Science, Rutgers University, 110 Frelinghuysen Road, Piscataway, NJ 08854, USA, snigdhamsushil@gmail.com}}
\affil{Rutgers University, Piscataway, NJ, 08854, USA}

\author{Laurent Burlion \footnote{Assistant Professor, Department of Mechanical and Aerospace Engineering, Rutgers Universoty, 98 Brett Road, Piscataway, NJ 08854, USA. AIAA Senior Member, laurent.burlion@rutgers.edu}}
\affil{Rutgers University, Piscataway, NJ, 08854, USA}

\begin{document}
\maketitle


%
\begin{abstract}
Sloshing Platform for In-Orbit Controller Experimentation is an ambitious, student run mission to design and fly a cubesat to study fluid sloshing in spacecraft. The project will examine zero-g propellant sloshing from an experimental standpoint. Despite the small size and limited payload capacity, we intend to use the cubesat platform to mimic larger spacecraft and implement novel detection and computer vision methods in our analysis. Many modern spacecraft rely on propellant-filled tanks to perform attitude control and station-keeping maneuvers. When a large percentage of the spacecraft's mass is comprised of liquid propellant, sloshing becomes a critical aspect of spacecraft attitude control and stability.  The mission will study the tank/fluid dynamics using new methods to gain an enhanced understanding of low-gravity fluid disturbance effects and improve simulations using equivalent mechanical models (EMMs). Active control of the fluid leading to the reduction of propellant slosh settling times will improve the maneuverability and performance of spacecraft. This paper will focus on satellite payload research and design requirements used to inform other aspects of the SPICEsat design. In this paper, mission objectives will be discussed, numerical simulations for the proposed control algorithms are demonstrated, and a satellite experiment design is presented. Finally, we examine computational fluid dynamics models to validate the satellite design and propellant sensing components of the proposed spacecraft.
\end{abstract}
\newpage
\section*{Nomenclature}
{\renewcommand\arraystretch{1.0}
\noindent\begin{longtable*}{@{}l @{\quad=\quad} l@{}}
$B_o$           & Bond Number - a dimensionless value measuring the gravitational versus surface tension forces \\
$I_{sat}$       & Moment of inertia of satellite \\
$\theta$        & Satellite angular position \\
$\Omega$        & Satellite angular velocity\\
$\Dot{\Omega}$  & Satellite angular acceleration\\
\modif{$\Omega_s$}      & \modif{Sloshing disturbance input to angular velocity}\\
\modif{$\Dot{\Omega}_s$}& \modif{Sloshing disturbance input to angular velocity dynamics}\\
$\Gamma_{RW}$   & Reaction wheel control torque \\
$\Gamma_s$      & Sloshing torque \\
$\dot\Gamma_s$  & Sloshing torque time dynamics \\
$\Gamma_{s_{NN}}$ & Sloshing torque neural network prediction\\
$\Gamma_d$      & External disturbances (i.e. atmospheric drag in low earth orbit) \\
\modif{$\Dot{\Gamma}_d$}& \modif{External torque disturbance dynamics} \\
\modif{$\Gamma_D$ }     & \modif{Sum of all external disturbances and sloshing ($\Gamma_s$ + $\Gamma_d$)}\\
\modif{$x(t)$   }       & \modif{Satellite state space vector} \\
\modif{$x(t)_{EMM}$ }   & \modif{Equivalent Mechanical Models state space vector} \\
\modif{$A_s,B_s,C_s,K_s$}&\modif{Equivalent Mechanical Models dynamics coefficients} \\
\modif{$H(S)$  }        & \modif{Transfer function governing reaction wheel dynamics} \\
\modif{$\sigma_{\Omega}$} &\modif{Gyroscope measurement standard deviation} \\ 
\modif{$\sigma_v$}        &\modif{Gyroscope angular random walk ($\degree / \sqrt{\text{time}}$)} \\
\modif{$\sigma_u$}        &\modif{Gyroscope bias instability ($\degree / \text{time}^{3/2}$)}\\
\modif{$\Delta t$}        &\modif{Gyroscope sampling time}\\
\modif{$\Delta P$}        &\modif{Sloshing pressure sensor disturbance measure}\\
NN              & Neural Network \\
EMM             & Equivalent mechanical models for approximating fluid sloshing in a tank \\
CFD             & Computational fluid dynamics \\
MoI             & Moment of Inertia \\
SPICEsat        & Sloshing Platform for In-Orbit Controller Experimentation \\
MSS             & Motion Sensing Suite \\
LSS             & Liquid Sensing Suite \\
VSS             & Vision Sensing Suite \\
sFSS            & sensor Fusion Sensing Suite \\
ADCS            & Attitude Determination and Control System \\
DoF             & Degree of Freedom \\
CONOPS          & Concept of Operations \\
\multicolumn{2}{@{}l}{Subscripts}\\
RW              & reaction wheel\\
s               & sloshing

\end{longtable*}}
\section{Introduction}
\subsection{Background}
Accurate attitude, positioning, and control are critical aspects of successful spacecraft operations. Satellite maneuvering accuracy is subject to disturbances generated by unspent propellants in the satellite's fuel tanks during orbital maneuvers or desired attitude changes. Fine guidance, such as needed for telescope pointing over long exposures, and coarse guidance, required when moving a satellite from one orbit to another, are all subject to disturbances generated by unspent propellants in the satellite's fuel tanks. The James Webb Space Telescope, launched in December 2021, is a highly accurate astronomical instrument requiring position stability accuracy of 6 milliarcseconds (6/1,000ths of an arc second) or 1.67 millionths of a degree. A single arcsecond = $\frac{1}{3600}$ of a degree.  The telescope contains tanks filled with three different propellants. According to the James Webb Space Telescope User Documentation \footnote{JWST User Documentation (JDox) \url{https://jwst-docs.stsci.edu/jwst-general-support/jwst-observing-overheads-and-time-accounting-overview/jwst-slew-times-and-overheads\#gsc.tab=0}, accessed 06-22-2024}, the maneuver rate is determined in part by the need to keep pointing accuracy settling times balanced with the need to reach the new pointing as soon as possible. For large slews between 25 arcseconds and 3 arcminutes, the slew rate is slower than for shorter or longer slews, to avoid exciting slosh modes of the propellant in the tanks. The propellant slosh is projected to take more than 20 minutes to dampen in some cases. Another  example of sloshing disturbances occurred in the March 2007 SpaceX Falcon 1 second launch attempt \footnote{Demo Flight 2 Flight Review Update, 2007, \url{https://forum.nasaspaceflight.com/index.php?action=dlattach;topic=7169.0;attach=506980}, accessed 06-22-2024} , where a control failure occurred at T+5 min, during second stage burn due to propellant slosh initiated by stage 1/2 contact at staging. Falcon 1 did not use slosh baffles in the second-stage tanks, as simulations done prior to flight indicated slosh instability was a low risk. Given that in space there are no gust or buffet effects, the simulations did not take into account \modif{this} perturbation, which occurred due to the hard slew maneuver after stage separation. Similarly, the effect of slosh is noted as a contributing factor in the \modif{Near Earth Asteroid Rendezvous} (NEAR) burn anomaly in December 1998, where a rendezvous burn resulted in the near loss of the spacecraft \footnote{The NEAR Rendezvous Burn Anomaly of December 1998, Technical Report, Applied Physics Laboratory, Johns Hopkins University, 1999, Hoffman, E.J. et. al.,\url{https://spacese.spacegrant.org/Failure Reports/NEAR_Rendezvous_BurnMIB.pdf} , accessed 06-30-2024}. NEAR’s two oxidizer tanks did not have an internal propellant management device (baffles or diaphragms) for slosh mitigation. Subsequent slosh studies resulted in in-flight modifications to the spacecraft's control algorithms. The sloshing problem was visually captured on January 29, 1964, during a Saturn 1 mission (SA-5), showing exactly how a fuel behaves in microgravity immediately after engine shutdown.  Similar footage is also available from the SpaceX CRS-4 Launch internal fuel tank camera showing fuel sloshing under orbital acceleration and immediately after engine shutdown. The dynamics are clearly complex.

Controlling fuel sloshing has been an engineering design concern since the dawn of the space age. In partially filled fuel tanks, large quantities of fuel can move inside the tanks under translational (thruster) and rotational (reaction wheels) accelerations. The fuel slosh develops from the dynamics of a free liquid surface inside of a fuel tank. As a spacecraft performs orbital maneuvers or attitude corrections, the free-moving propellant in a partially filled tank can excite low-frequency oscillatory modes as the free liquid surface undergoes complex motions imparting forces and moments on the vehicle. As noted by Bourdelle \cite{bourdelle2019a}, fuel mass as a percentage of total spacecraft mass can be considerable, and therefore mitigating sloshing effects through proper control algorithms becomes an important aspect of spacecraft design.

\subsection{Fuel Slosh Research}

On February 12th, 2005, the Dutch National Aerospace Laboratory (NLR) launched a satellite to examine the influence of liquid behavior on the dynamics of spacecraft. The SloshSat FLEVO (Facility for Liquid Experimentation and Verification in Orbit) experiment consisted of a cooperation with \modif{the European Space Agency} (ESA), the National Aeronautics and Space Administration (NASA) and a number of European subcontractors \cite{Vreeburg2005}. \modif{SloshSat included a capacitive liquid sensing system and thermal anemometers used to discern liquid flow}. Unfortunately, the satellite experienced problems resulting in the lack of any data on the condition of the water in the tank. As a result, all experimental results were derived from the accelerometer and gyroscope measurements of the motion of the satellite. To better understand the problem, in 2015 NASA, the Florida Institute of Technology (FIT), and the Massachusetts Institute of Technology (MIT) performed a series of slosh dynamics experiments on the International Space Station (ISS) using the Synchronized Position Hold, Engage, Reorient, Experimental Satellites (SPHERES) platform \footnote{Results of Microgravity Fluid Dynamics Captured With the Spheres-Slosh Experiment, 2015, Lapilli et. al, \url{https://ntrs.nasa.gov/citations/20150023503}, accessed 06-30-2024}.  These experiments were dedicated to studying the behavior of fluids in microgravity conditions, helping to validate, adjust, and improve Computational Fluid Dynamcis (CFD) models. \modif{SPHERES included the use of cameras to record fluid dynamics and the disturbance was measured using the experiment's inertial measurement units. Study of active control algorithms was not included in that experiment}.

In 2017, the French Space Agency, CNES (National Centre for Space Studies), conducted fluid sloshing experiments aboard the ISS. Named FLUIDICS (FLUIdDynamICsinSpace) \cite{FLUIDICS}, it was constructed like a slowly rotating centrifuge with two spherical tanks inside, and was capable of reproducing satellite sloshing behavior created when slewing a satellite. Sensor and visual data were collected by varying the fill ratios (two different tanks) and angular velocities. Forces and torques were measured from the experiment and compared against numerical simulations based on the traditional Navier-Stokes equations in zero-g \cite{FLUIDICS}. This research validated a number of assumptions that can be used in equivalent mechanical models, such as sloshing frequency and the effect of angular acceleration on slosh torques. 

Table \ref{table:experiment_compare} summarizes the various experiments conducted to examine the zero-g sloshing problem.

\begin{table}[H]
\centering
\begin{threeparttable}[b]
    \caption{\label{table:experiment_compare} Comparison of Zero Gravity Slosh Experiments}
    \renewcommand{\arraystretch}{0.85} 
    \begin{tabular}{cccccc}
        \hline\hline 
        \textbf{Mission} & \textbf{Force Sensor} & \textbf{Camera} & \textbf{Fluid Detection\tnote{1}} & \textbf{IMU Data \tnote{2}} & \textbf{Active Control} \\ 
        \hline 
            SloshSat     &      N               &       N   &           Y &                         Y &                 N \\
            SPHEREs      &      N               &       Y   &           N &                         Y &                 N \\
            FLUIDICS    &      Y               &       Y   &           N &                         Y (1-axis) &        N \\
            SPICEsat     &      Y (pressure)    &       Y   &           Y &                         Y &                 Y  \\
          \hline
    \end{tabular}
    \begin{tablenotes}
    \item [1] Fluid detection sensors (pressure sensors, capacitive liquid sensors, or thermal anemometers) 
    \item [2] Inertial Measurement Unit
    \end{tablenotes}
\end{threeparttable}
\end{table}
\renewcommand{\arraystretch}{1.0} 

Crosby \cite{Crosby2013} used a sounding rocket payload experiment to measure zero gravity fuel gauging using modal analysis. That experiment was successful in identifying three sloshing modes throughout the flight. In that research, Crosby also notes that zero-g parabolic flights with short periods of microgravity of approximately 20 seconds were insufficient to examine sloshing dynamics, as this does not allow enough time to measure the disturbance against the settled fluid. This drives experiments on sloshing to \modif{the} more stable and predictable zero-g environments of the ISS or a dedicated satellite.  More recent research by Vairamani and Crosby \cite{crosby2019} has taken the use of Propellant Management Devices from a passive instrument to an active one by using a magneto-active sloshing control system. Instead of rigid walls to contain the fluid, their research used a free floating membrane that becomes rigid when exposed to a magnetic field and therefore suppresses sloshing effects. Although ground based in 1g, the results do show a significantly shorter time for sloshing to settle in the tank when the active control mechanisms are employed.

For simulating sloshing behaviour,  the fuel sloshing problem has been studied numerically using computationally expensive CFD methods. CFD modeling is complex, especially \modif{when including dynamic} controls, where the inputs can change at each time step, introducing new dynamics into the problem and therefore requiring recalculation of the CFD solution. More commonly, the problem is studied using using Equivalent Mechanical Models (EMMs). These amount to approximating the fluid motion as a mechanical system such as a mass, spring damper or pendulum.  Smooth Particle Hydrodynamics (SPH) is also employed by some authors to approximate fluid dynamics. In SPH \cite{violeau}, the fluid is approximated as a collection of smaller elements (particles) that are each followed individually and their interactions are smoothed using a kernel function.

EMMs have been studied extensively in books and journal articles in such as Sidi \cite{sidi_1997}, Dodge \cite{dodge2000}, Ibrahim \cite{ibrahim2005} and  Reyhanoglu \cite{Reyhanoglu}. Peterson \cite{Peterson} studies the non-linear coupling between slosh modes and spacecraft dynamics in detail using a Lagrangian style EMM formulation. In their work, they noted that when coupled with the sloshing fluid, the spacecraft motion can exhibit strongly nonlinear motion. Attempting to predict spacecraft motion without coupling the interaction between the spacecraft and the sloshing fluid will predict motion different from experimental observations. Resonance motions also become a factor when coupling the two systems (spacecraft / slosh). 

Critical to the success of the EMMs is verification of the mechanical parameters (spring constants, mass, and pendulum length), which is typically accomplished using CFD models. For example, Enright and Wong \cite{cassini} evaluated the EMM parameters using CFD models for the Cassini spacecraft. In this research, we will use both EMMs and CFD models to approximate the sloshing behavior. The EMM work will be based on Bourdelle \cite{bourdelle2019b} where the physical parameters were modeled on the Detection of Electromagnetic Emissions Transmitted from Earthquake Regions (DEMETER) satellite. DEMETER was a French microsatellite mission launched in 2004 to observe geophysical parameters of Earth. CFD analyses are performed using Flow3D software from FlowScience \cite{FLOW-3D}.

\subsection{Paper structure}
This paper will focus on satellite payload research and design requirements used to inform other aspects of the SPICEsat design. Detailed designs for power consumption, total mass estimates, link budgets, pointing requirements, thermal control and other elements are available through the corrresponding author. The structure of this paper is as follows: Section I presents the propellant sloshing problem and its impact on spacecraft maneuvers. In Section II, we introduce the SPICEsat mission and discuss the proposed experiment detection methods, satellite actuation requirements, and experiment phases. In Section III, we introduce simulation work on various small satellite control methods and their application to SPICEsat. Section IV presents CFD models to further validate and refine the satellite design. Finally, in Section V, we present our conclusions.

\section{SPICEsat Mission Overview}

\subsection{SPICEsat Experimental NanoSatellite}

SPICEsat (Sloshing Platform for In-Orbit Controller Experimentation) will be a 6U satellite form factor, with one-third of the satellite consisting of a small, sealed tank filled with an appropriate propellant analogous liquid. Satellite construction started in early 2024 and is expected to be certified for a launch opportunity in 2025. Sloshing will be excited through a series of rotational maneuvers across all 3 axes to generate tank / fluid disturbance torques. We intend to explore mid-range Bond numbers $1 < B_o < 10$. The Bond number is a dimensionless number measuring the importance of gravitational forces compared to surface tension forces. Bond numbers of ten or less are generally considered to represent "reduced" gravity; therefore SPICEsat mission results will apply to satellites in micro-gravity conditions undergoing typical slew maneuvers. The mission will characterize these sloshing effects by measuring standard attitude parameters and applying existing techniques to the sloshing problem to gain new insight into the problem. Three methods of fluid characterization are to be explored: Motion sensing with accelerometers and gyroscopes, a pressure "sensor map" of fluid/tank interactions, and recorded video footage enhanced by computer vision analysis. 

\vspace{5mm}
\underline{SPICEsat will have 3 main objectives:}
\begin{enumerate} [leftmargin=45pt]
    \item Demonstrate the feasibility of using pressure sensor arrays to map tank/fluid dynamics
    \item Demonstrate the feasibility of using computer vision to analyze complex sloshing dynamics
    \item Demonstrate the slosh disturbances mitigation strategies using novel active control algorithms
\end{enumerate}
\vspace{5mm}

To accurately determine the spacecraft's attitude, the experiment will require high-precision gyroscopic measurements. An onboard motion sensing suite (MSS) is expected to measure rotational motion and disturbances. This sensor suite should improve upon the 2005 SloshSat FLEVO experiment. SloshSat suffered a mission anomaly immediately upon signal acquisition. Unfortunately, downlinked data did not include any in-tank experiment information. Therefore all results from SloshSat were derived from its accelerometers and gyroscopes. SloshSat FLEVO used early 2000s era technology e.g. Litef $\micro$FORS 36/6 \cite{LiteuFORS} fiber optic gyroscopes \big(Bias $\leq 18 \degree$/hr (1$\sigma)$ and Noise $\leq 1\degree/\sqrt{hr}$ \big). Modern gyroscopes such as the EPSON M-G364 \cite{EPSON}, are at least an order of magnitude more sensitive \big(Bias $\leq 2.2 \degree$/hr (1$\sigma)$ and Noise $\leq 0.09\degree/\sqrt{hr}$ \big). We will use modern, highly sensitive gyroscopes (sensitivity > 0.05 \degree/s) and accelerometers to measure changes in attitude and fluid forces on the spacecraft caused by sloshing in the tank. SPICEsat is expected to \modif{utilize reaction wheels capable of torques, $\Gamma_{RW}$, of up to $0.006$ N-m resulting} in estimated sloshing torques of $\Gamma_s$  $< 10^{-3}$ N-m, and translating into tangential accelerations \modif{of up to} 25 $\mu$g, which is above the detection threshold of modern accelerometers. \modif{Detailed analyses of detection thresholds is presented in Section IV}.   

A liquid sensing suite (LSS) will create a coarse "pressure map" of the actual forces the fluid imparts on the tank wall. Ideally, lining the entire tank would form a continuous sensing surface, but this precludes the recording of video. Therefore, the sensor array will consist of 8 strips, placed uniformly inside the tank. 

A video sensing suite (VSS) will record video of the sloshing effects and transmit this information to the ground for further processing. Lightweight and low-power cameras are being explored since the experiment only requires knowledge of the location of the fluid. Black-and-white, low frame-rate video footage will be used in order to reduce file sizes for later transmission to Earth. Modern computer vision techniques will be employed on the ground to visualize fluid/tank wall interactions \modif{and} fluid oscillations. 

The satellite is expected to conduct up to \modif{229} different experiments \modif{gathering} sloshing data for download to a ground station and further analysis and interpretation. Ground software will be created to analyze the MSS data and reconstruct the state space vector \modif{to compare against} existing Computational Fluid Dynamics (CFD) models, tune the CFD parameters representing the fluid, and further validate existing EMMs. Satellite state space, pressure, and vision data will be combined and used to train a machine learning (ML) model to predict sloshing effects. Each experiment will be repeated 3 times in reproducible data sets. Reaction wheel control inputs will be exercised along each axis of the satellite individually (x,y,z), then simultaneously along two axes (xy,xz,yz), and finally along all three axes at once (xyz), for a total of seven different excitation modes. Each experiment will begin by using a baseline control algorithm with no slosh mitigation strategies enabled, then a post-excitation mitigation strategy to dampen the slosh effect. Once sufficient data is collected and analyzed, the satellite will be programmed in flight to implement four \modif{different} control algorithms. A baseline, standard Attitude Determination and Control System (ADCS) controller will be used and will also act as the safe-mode controller for the remainder of the experiment. Next, a time delay \modif{adaptive} control algorithm (also called an Output Feedback \modif{Adaptive} Controller) will be implemented, where the controller uses a series of past values and implements a control strategy designed to mitigate sloshing. Then, a ground-trained ML controller "black box" will be uploaded to the satellite, and a control solution implemented using sloshing predictions which will be tested to reject the sloshing effects. \modif{Next, a "Reference Governor" style controller will be implemented. Reference Governors are used to ensure system dynamics respect a given set of constraints. Finally, the ML and Output Feedback Adaptive controllers experimental data will be compared and the one minimizing the settling time will be combined with the reference governor} and tested. Significant work has already been accomplished validating the EMMs \cite{bourdelle2019a},\cite{bourdelle2019b}, and \cite{bourdelle2019}, and applying machine learning to satellite state vectors for predictions \cite{fogel2022}. This ML approach and a description of the numerical results are highlighted in Section III.

\subsection{Experiment Equations of Motion and Flight Requirements}
The basic equation of motion governing this problem is well known. Consider a spacecraft rotating rigid body in the absence of any disturbances:

\begin{equation}
    I_{sat}\Dot{\Omega} = \Gamma_{RW}
    \label{IsatOmega=Trw}
\end{equation}

where $I_{sat}$ is the moment of inertia (MoI) of the satellite (single axis of rotation), $\Dot{\Omega}$ is the angular acceleration of the satellite, and $\Gamma_{RW}$ is the torque generated by the reaction wheel. Taking sloshing and other disturbances (i.e. radiation pressure or Low Earth Orbit (LOE) atmospheric drag) into account, the equation can be re-written as:

\begin{equation}
    I_{sat}\Dot{\Omega} = \Gamma_{RW} + \Gamma_s + \Gamma_d
        \label{IsatOmega=Trw+AllDisturbances}
\end{equation}

where $\Gamma_s$ is the sloshing torque and $\Gamma_d$ represents any other external disturbances. In Table \ref{table:sloshing_vars}, we define each term :

\begin{table}[ht]
  \centering 
  \caption{Satellite Sloshing Parameter Analysis} 
  \begin{tabular}{c c l}
    \hline\hline 
    \textbf{Parameter} & \textbf{Measure} & \textbf{Description}  \\ 
    \hline 
    $I_{sat}$           & Known       & MoI matrix measured pre-flight \\
    $\Dot{\Omega}$      & Measured    & Satellite angular acceleration measured in-flight\\
    $\Gamma_{RW}$       & Known       & Commanded control torque \\
    $\Gamma_s$          & Unknown     & Sloshing torque \\
    $\Gamma_d$          & Unknown     & External disturbances (i.e. atmospheric drag in LOE) \\
    \hline 
  \end{tabular}
  \label{table:sloshing_vars}
\end{table}

 Since $\Gamma_s$ and $\Gamma_d$ cannot be easily distinguished from one another, this immediately places a number of requirements on SPICEsat that must be considered. First, if we consider the entire system as a closed-loop control system, we can generate a reaction wheel torque that compensates for both, such that:
 
\begin{equation}
    I_{sat}\Dot{\Omega} = \Gamma_{RW} + \Gamma_D
        \label{IsatOmega=Trw+Disturbances}
\end{equation}

where

\begin{equation}
    \Gamma_D = \Gamma_d + \Gamma_s
    \label{T_Disturbances}
\end{equation}

For the duration of each experiment, we assume:
\begin{equation}
    \Dot\Gamma_d = 0
    \label{T_disturbances=0}
\end{equation}

Although $\Gamma_d$ is uncertain, measurements of atmospheric drag in LOE are expected to be small, between \num{1e-8} km/s$^2$ and \num{1e-10} km/s$^2$ and we are assuming them to be constant during the periods of our experiment, typically lasting 100 - 500 sec. Using the additional measurement methods proposed in this document, we aim to bound the interval of uncertainty in measuring $\Gamma_s$.

\subsection{Experiment Overview} 
\begin{figure}[H]
    \centering
	\includegraphics[width=0.60\textwidth]{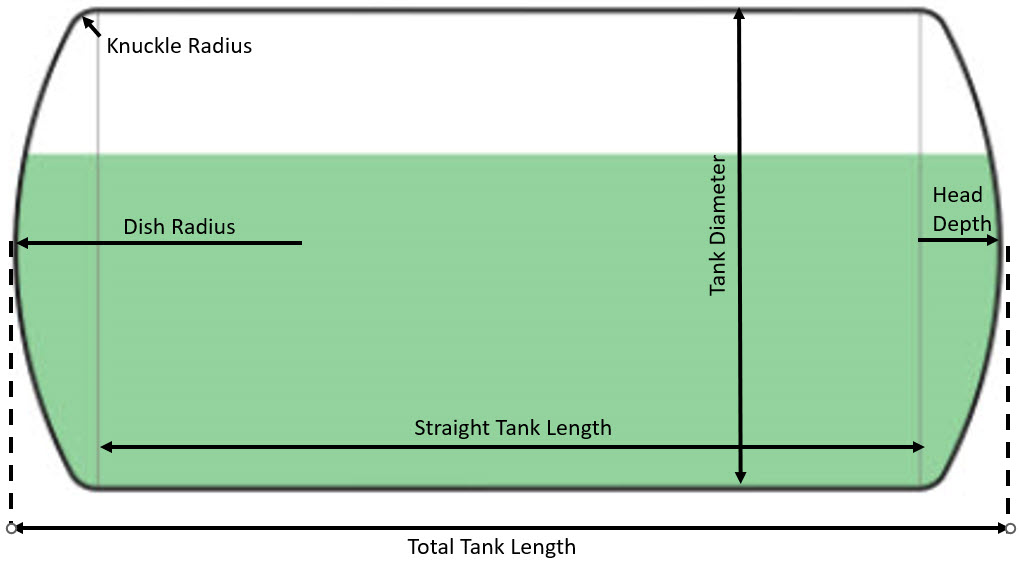}
	\caption{SPICEsat sloshing tank geometry visual overview showing key parameters and liquid fill level.}
	\label{fig:tank_geometry}
\end{figure}

SPICEsat's primary experiment centers around a transparent, sealed tank containing a liquid analogous to spacecraft propellant, specifically, deionized water. To generate the necessary sloshing excitations with a Bond Number, $1 < B_o < 10$, the characteristic length scale of the tank is required to be sufficiently long. Therefore to extend the length of the tank, SPICEsat must be a 6U satellite. A schematic representation of the proposed tank is illustrated in Fig \ref{fig:tank_geometry} with other relevant physical parameters of the tank are shown in Table \ref{table:tank_params}. The tank is expected to have a 90 mm diameter and a length of 160 mm for a total volume of 0.942 liters (0.000942 $m^3$) and a fill ratio of 70\% or 0.709 liters (0.000709 $m^3$). Assuming a total 6U satellite mass of 7 kg and distilled water as the fluid, fluid mass represents just under 13\% of the total SPICEsat mass.

\begin{table}[H]
\centering
  \begin{threeparttable}[t]
    \caption{Slosh Tank Physical Parameters}
    \label{table:tank_params}
    \renewcommand{\arraystretch}{0.85} 
    \begin{tabular}{ll}
    \hline\hline 
    \textbf{Parameter} & \textbf{Measure}  \\ 
    \hline
            Orientation             & Horizontal    \\
            Vessel Head             & AMSE Flanged \& Depth \tnote{1} \\ 
            Shape                   & Torispherical \\
            Diameter                & 90.00 mm      \\
            Vessel Head Depth       & 15.24 mm      \\
            Straight Length         & 129.5 mm      \\
            Total Tank Length       & 160 mm        \\
            Dish Radius Parameter   & 1.00          \\
            Knuckle Radius Parameter& 0.06          \\
            Head Type               & Convex        \\
        \hline
     \end{tabular}
     \begin{tablenotes}
       \item [1] American Society of Mechanical Engineers standard.
     \end{tablenotes}
  \end{threeparttable}
\end{table}
\renewcommand{\arraystretch}{1.0} 

\subsection{Sensing Suites} 
The satellite will be instrumented with three different measurement suites and a ground-based sensor fusion algorithm. Each suite will be developed to facilitate the detection, data collection, and transmission of the sloshing effects on the attitude of the spacecraft.

\begin{enumerate}
    \item MSS - Motion Sensing Suite
    \item VSS - Vision Sensing Suite
    \item LSS - Liquid Sensing Suite
    \item sFSS - sensor Fusing Sensing Suite (ground) 
\end{enumerate}

\subsubsection{Motion Sensing Suite (MSS)}
\begin{figure}[H]
    \centering
	\includegraphics[width=0.60\textwidth]{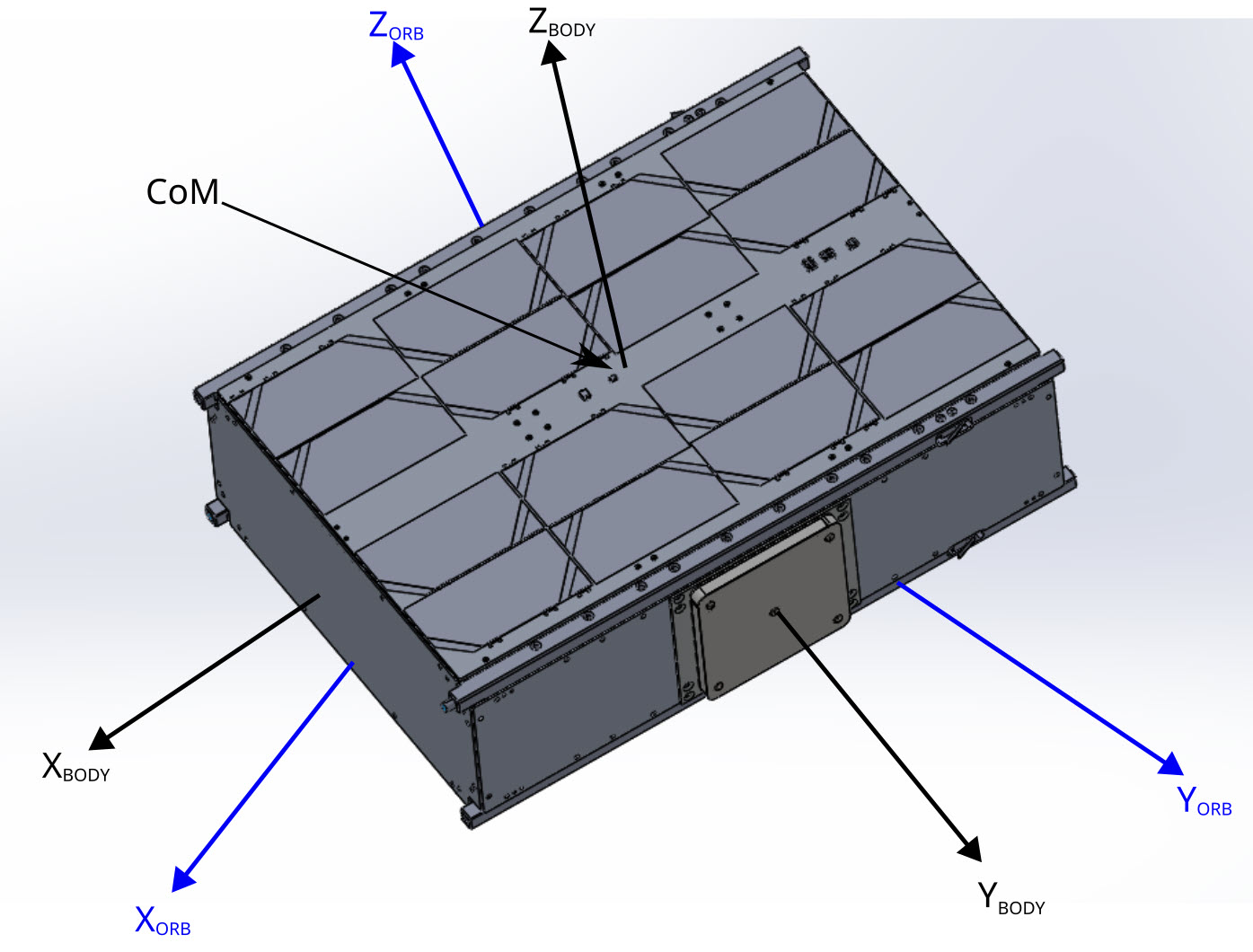}
	\caption{6U satellite coordinate system diagram, inspired by \cite{CAT-2}}
	\label{fig:MOI-Coordinates}
\end{figure}

SPICEsat's filled tank, the heaviest component, will be positioned at the end of the 6U nanosatellite form factor. Because we expect to be taking video images of the fluid slosh, a large field of view is needed for the camera to see the tank. Therefore, the center portion of the spacecraft will only contain one significant component. Balancing the tank at the opposite end of the spacecraft will be the ADCS, computers, and communications components. This very effectively positions the center of mass near the geometric center of the spacecraft, as depicted in Fig \ref{fig:MOI-Coordinates}. The MSS will consist of high-precision accelerometers and gyroscopes, included in the spacecraft ADCS package. Sloshing is well known to be a relatively slow-moving disturbance. Research in the field by Green \cite{slosh_freq} indicates sloshing frequencies $\approx 1 $Hz are typical. Sloshing equivalent mechanical models \modif{agree with those estimates, therefore}, a minimum sensor time resolution of 5 Hz or better will be required. 

The SPICEsat team has used a Computer Aided Design (CAD) model to determine the moment of inertia on of $I_{sat} = 0.0542 $ kg-m$^2$ in a body coordinate system as shown in Fig. \ref{fig:MOI-Coordinates}. The SPICEsat coordinate system and moment of inertia model are consistent with a typical 6U satellite model. For a similar example, see CAT-2\cite{CAT-2}. 

The equivalent mechanical models indicate a sloshing torque of \num{1.0e-4} $< \Gamma_s < $ \num{1.0e-3} N-m. In Section IV, we model the sloshing torque using CFD calculations for the same size tank and reaction wheel torque studied in the EMMs. The CFD models show similar sloshing torques of \num{1.4e-3} $< \Gamma_s < $ \num{7.6e-3} N-m. Assuming the spacecraft center of mass is located 10 cm from the center of the tank (in a 6U satellite), we can calculate the angular and linear accelerations as: 

\begin{equation}
    \dot\Omega_{min} = \frac{\Gamma_s}{I_{sat}} = \frac{0.000134}{0.0542} = \num{2.47e-3} \text{rads/}s^2
    \label{Omega_dot(min)}
\end{equation}

\begin{equation}
    a_{min} = \dot\Omega_{min} r = (\num{2.47e-3})(0.1) = \num{2.47e-4} \text{m/}s^2 = 25.2 \micro g
        \label{linear_accell(min)}
\end{equation}

The sloshing torques were taken from EMM (Section III.A) and CFD (Section IV.D) models to estimate the order of magnitude of the angular acceleration ($\dot\Omega_s$) and angular velocity ($\Omega_s$) disturbances caused by the sloshing ADCS maximum reaction wheel torque of 0.006 N-m (a BlueCanyon XACT-50). This will give us an order of magnitude measurement of the effect of sloshing, which we will require the gyroscopes to detect. 

The results show that the typical angular acceleration disturbance we can expect from sloshing to be $1.5 \degree/s^2 < \dot\Omega_s < 7.8 \degree/s^2$. If we assume a gyroscope sampling rate at 10 Hz, this gives a gyroscope measurement sensitivity requirement @10 Hz of $0.15 \degree/s < \Omega_s < 0.78 \degree/s$.

Validation work is presented in Section IV to further these ADCS requirements and sensitivities using CFD software. Center of Mass, fluid pressure, force, and velocity fields are calculated and will be used to ensure satellite components selected for the mission are suitable.

To estimate MSS data requirements, we expect to get seven data fields from the ADCS (three gyroscopic, three accelerometers, and one timestamp). The internal format of a timestamp is a seven to thirteen-byte composite value \footnote{IBM Corporation, "Date, Time, and Timestamps",2022,\url{https://www.ibm.com/docs/en/i/7.3?topic=concepts-date-time-timestamps},accessed: 03-05-2022} and is estimated to be 80 bits measured at 100 Hz:

\vspace{-0.5cm}
\begin{equation}
    \textbf{TimeStamp Data (ms)} = \text{80 bits x 100 Hz = 8000 bits/sec} 
\end{equation}

For the gyroscope and accelerometers, a double precision floating point number requires 8 bytes of storage or 64 bits. Assuming this, we have

\vspace{-0.5cm}
\begin{equation}
    \textbf{MSS Measurement bits} = \text{100 Hz x 6 data x  64 bits/data = 38,400 bits/sec}
\end{equation}

\vspace{-1cm}
\begin{equation}
    \textbf{Total MSS bits = Data + TimeStamp} = \text{38,400 + 8000 = 46,400 bits/sec} 
\end{equation}

We expect each sloshing and mitigation to take between 100 - 500 seconds to settle and end the experiment. Therefore the total MSS data requirements in Megabyes (MB) and Gigabytes (GB) become:

\vspace{-0.5cm}
\begin{equation}
    \textbf{MSS Data} = \text{46,400 bits/sec x 500 sec = \num{23.2e+6} = 2.9 MB/experiment}
\end{equation}

\subsubsection{Vision Sensing Suite (VSS)} 
\begin{figure}[H] 
\centering
\renewcommand*{\arraystretch}{0}
\begin{tabular}{*{3}{@{}c}@{}}
(a)  & (b) & (c) \\
\includegraphics[scale=0.40]{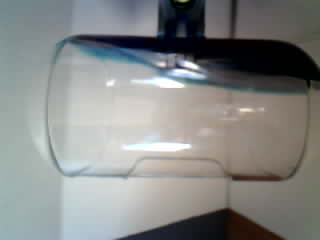}    & 
\includegraphics[scale=0.58]{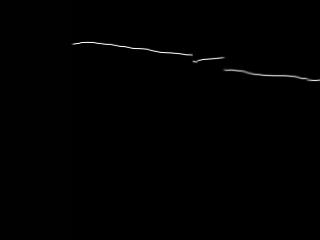}   & 
\includegraphics[scale=0.58]{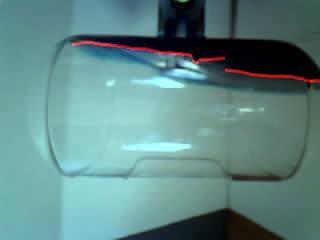}    \\
\includegraphics[scale=0.40]{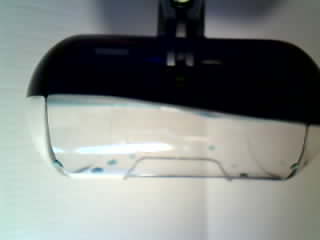}   & 
\includegraphics[scale=0.58]{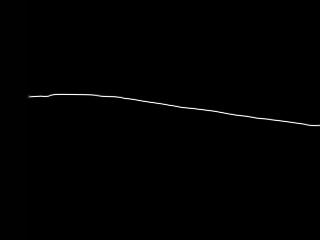}  & 
\includegraphics[scale=0.58]{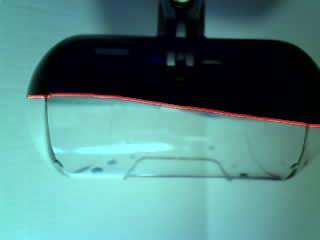}    \\
\includegraphics[scale=0.40]{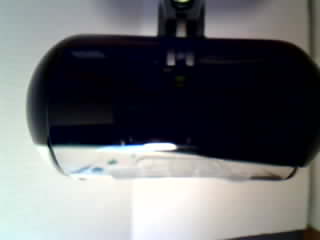}  &
\includegraphics[scale=0.58]{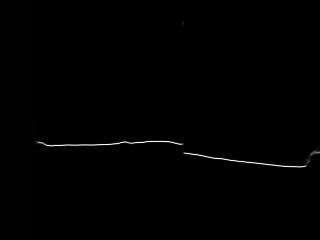} &
\includegraphics[scale=0.58]{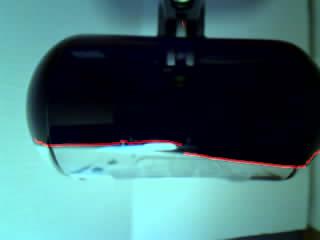} \\   
\end{tabular}
\caption{Sloshing images taken from a testbench constructed to represent the camera/tank positioning, angle, and camera field of view. Column (a) shows input image samples. Column (b) shows the predictions of the CNN model (white pixels in these images represent liquid boundary predictions.) Column (c) shows the boundary predictions of Columns (a) and (b) superimposed.}
\label{fig:1g_slosh_CNN}
\end{figure}

The VSS is a mid-resolution, low frame-rate, black-and-white camera designed to capture the fluid motion inside the transparent tank. The choice of resolution and black-and-white video is driven by the desire to reduce data sent to the ground station. Additionally, color video adds very little information to the problem. Several lightweight and small form-factor cameras are being evaluated, such as an Arducam OV5647. \modif{ Ground processing software is required to analyze the VSS video using convolutional neural networks. The Burlion Advanced Controls Lab at Rutgers University has successfully conducted preliminary work using public-domain software to detect and measure the surface disturbance of a sloshing liquid in 1g. For this purpose, a three layer Convolutional Neural Network (CNN) was designed and trained to detect the top line of the sloshing/air interface in the transparent tank. The foreground boundary of the sloshing/air interface was selected as the relevant liquid boundary for analysis. This neural network was built using PyTorch \cite{paszke2019pytorch}, an open-source tensor library for machine learning. A 1 Degree of Freedom (DoF), 1-g rotating testbench has been designed to determine the effectiveness of the proposed CNN algorithms. The testbench is constructed to represent the camera/tank positioning, angle, and camera field of view. The CNN model was trained using fourteen images from the testbench. A validation set of three additional images was used to select the best model and training parameters. Fig. \ref{fig:1g_slosh_CNN} shows the output of the CNN model trained for 1000 iterations using \textit{Adam} \cite{KingmaB14}, a gradient descent-based optimizer. In Fig \ref{fig:1g_slosh_CNN}, Column (a) shows input image samples that were excluded from the model training and validation data. Column (b) shows the grayscale images of predictions of the CNN model. White pixels in these images represent liquid boundary predictions. Column (c) shows the boundary predictions superimposed with the original images. Boundary prediction lines are in red, and have been shown with exaggerated width for clarity. The top and bottom row image samples have significantly different liquid volumes, compared to the model training images. We see that the CNN model's accuracy is largely invariant to liquid volume, illumination changes and orientation. It should be noted that the camera captured these images upside down, however the CNN trained model can be trivially extended for prediction on upright images. 
\\
This method can be used to measure and estimate the sloshing frequency needed as input to most EMMs. The software will be improved to detect interactions between the fluid and the side of the tank, as well as the approximate location of that impact. For zero-g experiments, we intend on using fluid bubble tracking Computer Vision techniques to track fluid/tank wall interactions, bubble size, velocities, and directions and estimate momentum imparted into the tank. Tying this visual information with pressure sensor data and MSS measurements should permit sloshing identification. }

To calculate video storage and transmission requirements, we use the following input parameters. These conservative estimates will need to be refined based on the camera selection, resolution, and frame rate as specified in Table \ref{table:video_params}.

\begin{table}[H]
    \caption{\label{table:video_params} VSS Camera Parameters }
    \centering
    \renewcommand{\arraystretch}{0.85} 
    \begin{tabular}{ll}
        \hline\hline 
        \textbf{Parameter} & \textbf{Measure}  \\ 
        \hline 
            No. of Cameras           & 1            \\
            Resolution              & 960H x 540V  \\ 
            Recording Time          & 500 sec      \\
            Frame/Sec               & 24 fps       \\
            No. of Experiments      & 1            \\ 
            Compression             & H.264        \\
        \hline
    \end{tabular}
\end{table}
\renewcommand{\arraystretch}{1.0} 

The resulting data file is close to 104 MB per experiment for image data. 

\vspace{-0.5cm}
\begin{equation}
    \textbf{Vision Sensing Suite} =  \text{ 104 MB / experiment}
\end{equation}

\subsubsection{Liquid Sensing Suite (LSS)}
A third proposed method of detecting sloshing within a propellant tank is the use of pressure sensors to measure a quantity proportional to the sloshing torque. \modif{Specifically, the pressure sensor pads will measure the force exerted when the floating liquid interacts with the wall of the tank at that specific position.} Then, using this measurement, an attitude controller can be designed not only to reject the sloshing torque but also to keep the fuel sloshing within acceptable conditions. This constraint condition is achieved by combining model-free control with an output-to-input saturation transformation (OIST) add-on. A first attempt to use pressure sensors in a tank is researched by Zhao, Fogel, and Burlion \cite{zhao} using a sloshing equivalent mechanical model (EMM). The authors have shown that using a standard PID controller and adding a pressure measurement constraint to that controller can limit the sloshing effect to within a predetermined min/max. \modif{The pressure sensor measurements will also be used in a reference governor technique \cite{Kolmanovsky-Tutorial} as discussed in Section II.F.}  

\begin{figure}[H]
    \centering
    \includegraphics[width=11cm]{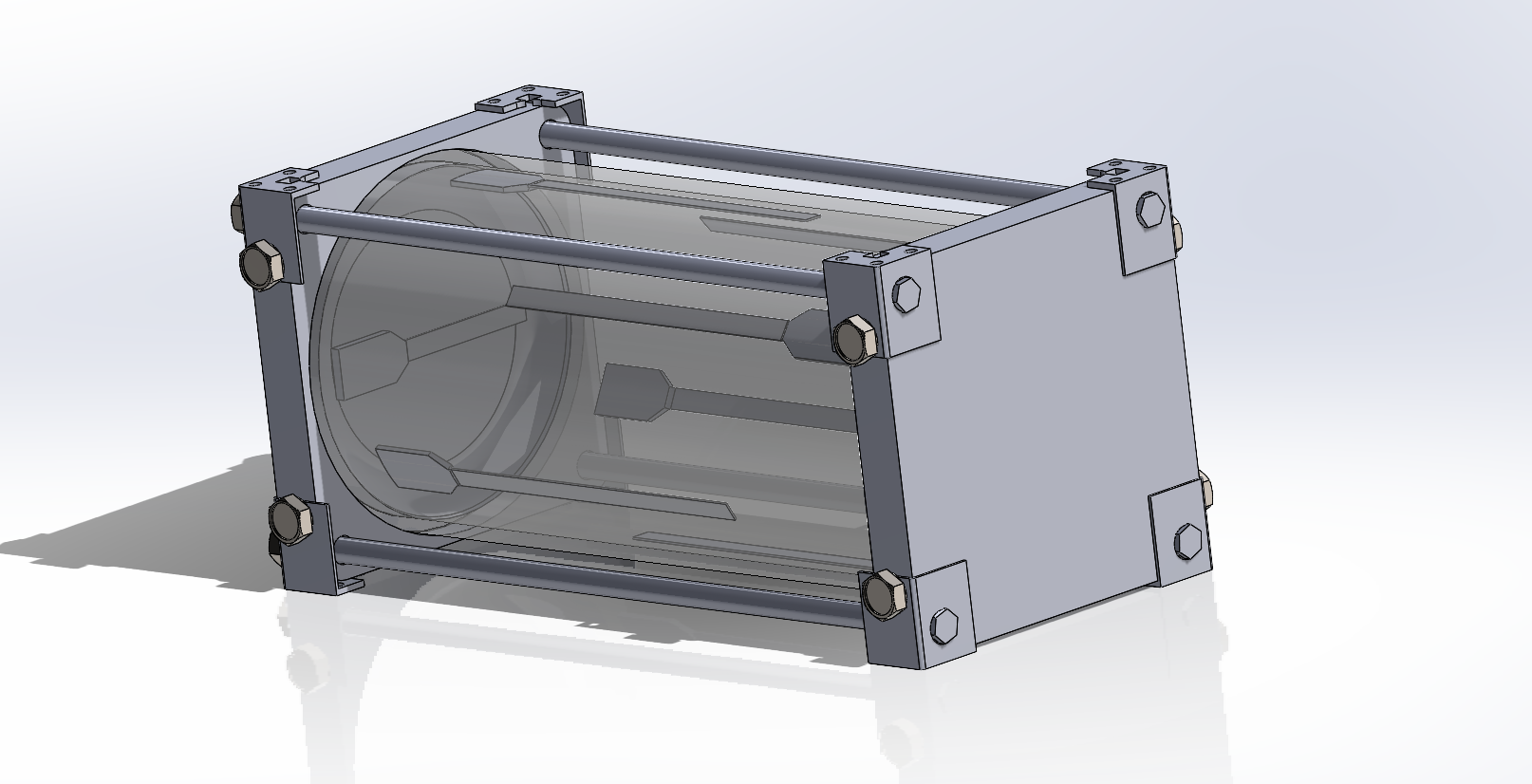}
    \caption{Illustration of pressure sensor array inside tank.}
    \label{fig:sensorarray}
\end{figure}

SPICEsat will be instrumented with pressure sensors located uniformly throughout the tank. Fig. \ref{fig:sensorarray} shows a schematic illustration of the proposed sensors inside the tank.

This figure illustrates a coarse pressure sensor array of at least eight sensors total, with six sensors lining the inside wall of the tank and one sensor at each end. Inserted horizontally, the sensors will divide the tank into six sectors and one sensor at each tank end cap. This configuration allows for vertical and horizontal measurements of the tank-fluid interaction. Sensor locations were determined to obtain a reasonable distribution of detection area while still allowing video recording of fluid movement.  Initially, capacitive pressure sensors were considered. This option was eliminated due to their need to equalize pressure external to the sensor itself via a pressure relief tube. New custom sensors are being investigated from Sensitronics, which consists of 16 sensor pads in a single strip. Each pad measures 5 mm x 5 mm and is the limiting resolution. The sloshing motion will be captured by the sensor pad and \modif{will} transmit two bytes per array sensor per frame, plus a single terminator byte to signal the end of a frame/beginning of the next frame. The sensors will be calibrated using a per-sensor tare in-situ rather than applying a simplistic global “noise floor” because a “noise floor” hides the low-end sensitivity of many of the individual sensor locations by applying an excessive threshold. For sensors pads that have higher preloads, the sensors are expected to have an activation in the 0.01-0.02 psi range which will be validated in the lab before installation into the satellite.

\vspace{-0.5cm}
\begin{equation}
    \textbf{Pressure Sensor \modif{Sensitivity}} = \text{0.01 - 0.02 psi}
\end{equation}

Sensitronics expects to have approximately 12 usable bits of raw sample resolution. With N strips of sixteen sensor pads each, then each frame would be (N strips x 16 sensors x 2 Bytes/frame) + 1 terminator byte = (N x 32 Bytes/frame) + 1 Byte/frame. Assuming N = 8 at a frame rate of 10 frames/sec (10 Hz): 

\vspace{-0.5cm}
\begin{equation}
    \textbf{Pressure Sensor Data} = \text{32 Bytes/array/frame x 8 strips} + \text{1 Byte} = \text{ 257 B / frame}
\end{equation}

\vspace{-1cm}
\begin{equation}
    \textbf{Pressure Sensor Data} = \text{257 Bytes/frame x 10 frames/s} = \text{ 2,570 Bytes/s} = \text{ 0.0257 MB/s}
\end{equation}

The pressure sensors will be connected in a “daisy chain” configuration with associated conditioning electronics that output standard Serial Peripheral Interface (SPI) protocol and will communicate with the onboard payload computer. The flight computer will then handle any additional data processing and storage required prior to downlink. 

\vspace{-0.5cm}
\begin{equation}
    \textbf{Pressure Data/Experiment} \\
    = \text{ 0.0257 MB/s x 500 s/exp} = \text{ 12.85 MB/exp}
\end{equation}

\subsubsection{Total Data Requirements}

During the lifetime of the experiment, we expect to conduct up to \modif{229} experiments, including three trials of each experiment, seven different rotations, and three satellite modes across four different controller types. One-third of those experiments \modif{(76)} will include camera data, and the remaining two-thirds \modif{(153)} experiments will include sensor data only. The MSS, VSS, and LSS all contribute to the total experiment data download requirement. 
The satellite itself also transmits important State of Health (SoH) data to the ground in addition to the Experiment Data. This includes information such as battery charge over time, satellite temperature, ADCS statistics, etc. This SoH data is designed to ensure the non-payload related components are functioning properly. Therefore, in total, the data download requirements include both satellite State of Health (SoH) and Experiment information and are summarized in Table \ref{table:data_require_summary}.

\begin{table}[H]
    \caption{\label{table:data_require_summary} SPICEsat Data Requirements (MB)}
    \centering
    \renewcommand{\arraystretch}{0.85} 
    \begin{tabular}{clllll}
        \hline\hline 
        \textbf{ } & \textbf{VSS} & \textbf{LSS} & \textbf{MSS} & \textbf{SoH} & \textbf{Total} \\ 
        \hline 
            With Camera         & 104 & 12.85   & 2.9 & 7 & 126.75 \\
            Without Camera      & N/A   & 12.85   & 2.9 & 7 & 22.75 \\ 
          \hline
    \end{tabular}
\end{table}
\renewcommand{\arraystretch}{1.0} 

The total payload data requirements throughout the mission are:

\begin{center}
  \begin{center}
    \textbf{76  Experiments @ 126.75  MB}  = 9,633 MB \\
    \textbf{153 Experiments @ 22.75 MB}    =   3,481 MB \\
    \textbf{   ALL experiments: 229 Experiments} = 13,114 MB \\
    \vspace{2.5mm}
  \end{center}
\end{center}
\vspace{-1cm}

\subsubsection{Sensor Fusion Sensing Suite (sFSS)}
All sensor data is expected to be recorded, transmitted to the ground, and analyzed. All data should be time-stamped and synchronized on-board SPICEsat with the intention of comparing satellite motion (MSS), the pressure measurements (LSS) by location within the tank, and the corresponding visual interaction (VSS) of the fluid with the side of the tank. By fusing all the data into a single analysis, we intend to recover full satellite state space (Eq. \ref{extended_state_space}).

\begin{equation}
x_{EMM}(t) \\
= \begin{bmatrix}
        \theta(t)    &
        \Omega(t)      &
        \dot{\Omega}(t)  &
        \Gamma_s(t)    &
        \dot\Gamma_s(t) 
    \label{extended_state_space}
\end{bmatrix}^T
\end{equation}

Once the state vector is recovered from the sensors, the data can be used to validate the existing EMMs and used to re-design and upload new control strategies to SPICEsat.

\subsection{Experiment Operating Modes and Design} 

\subsubsection{Slosh Excitation}
The scientific portion of this project requires SPICEsat to maneuver in various directions and attitudes to generate sloshing modes inside the cylindrical tank. Understanding sloshing characteristics of maneuvers that mimic larger satellite flight characteristics is critical. This will require a set of pre-programmed maneuvers for SPICEsat similar in nature to the SloshSat experiment \cite{Vreeburg2005}. The satellite will cycle through each of the maneuvers for every new controller algorithm tested.  Each experiment should \modif{start from a pre-established initial condition} and conclude when the fluid is at rest again. Subsequent experiments should only begin after the fluid is completely settled.

\vspace{0.25cm}
\begin{enumerate}
    \item \textbf{Rotation About a Single Axis} – Spin the tank at a specific angular rotation rate along on single axis (x,y,z)
    \item \textbf{Rotation About Two Axes} – Spin the tank about two axes (xy, yz, xz)
    \item \textbf{Rotation About Multiple Axes} – Spin the satellite about multiple axes (xyz) simulating a tumble
\end{enumerate}

To ensure adequate control authority throughout each experiment, we have conducted both EMM and CFD calculations using Flow-3D to determine the total forces and torques the sloshing fluid will exert on the satellite. The EMMs indicate a sloshing torque on the order of 0.001 N-m. The currently selected ADCS has a maximum control torque of 0.006 N-m. Flow-3D model validates this approach, showing a sloshing torque of 0.001 N-m to 0.008 N-m.

\subsubsection{Experiment Operating Modes} 

The experiment will require multiple satellite operating modes to accommodate science objectives.

\vspace{0.25cm}
\begin{enumerate}
    \item \textbf{Excitation \& \modif{Observation}} – Excite sloshing using reaction wheels. \modif{Enable all sensor suites, observe, and record angular rate, angular acceleration, pressure sensor, and video data.}
     \item \textbf{Excitation \& Mitigation} – Controlled excitation of sloshing while simultaneously using \modif{newly developed} control algorithms to mitigate slosh
    \item \textbf{Idle} – All maneuvers and observations complete, fluid settles until next experiment
    \item \textbf{Recovery} - Safe mode controller to bring the satellite and fluid back under control in the event of excess excitation
    \item \textbf{Algorithm Update} – Enable the ADCS to accept a newly uploaded control algorithm
\end{enumerate}\vspace{0.25cm}

\subsection{Slosh Mitigation Algorithms}

After a period of data gathering and sensor analysis on the ground using system identification techniques like SINDy (Sparse Identification of Nonlinear Dynamics) \cite{SINDyBrunton} and DMD (Dynamic Mode Decomposition) \cite{DMDproctor2014dynamic}, various control algorithms will be explored to mitigate slosh effects on the satellite. A primary output of this data collection and analysis is to iterate through various control algorithms using ground simulations. This goal is to maximize slosh mitigation performance. We will have \modif{three} methods to test these controllers, 1) Numerical simulations using EMMs and 2) a 1-g, 1 Degree of Freedom (DoF) test bench experiment\modif{, and 3) "CFD in the loop" simulations} prior to uploading to the satellite. 

The data used from the satellite can be used to train a machine learning (ML) algorithm directly. Using the MSS data and the known initial control input, an ML model can be trained to predict the sloshing effect and therefore compensate for it in the control feedback loop. This is discussed in more detail in Section III, \modif{where we present results testing various neural network algorithms}. This iterative approach to satellite control requires the flight control software to have the capability to upload new algorithms while always preserving a standard controller in case of a recovery situation. 
\modif{The following four different control algorithms will be implemented:}
\vspace{0.25cm}
\begin{enumerate}[label=]
    \item \textbf{A. Baseline Controller} – \modif{Commercial off-the-shelf control algorithms in the onboard ADCS.}  
    \item \textbf{B. Output Feedback \modif{Adaptive} Controller} – Uses past sensor and state vector values to modify an existing Proportional, Integral, Derivative (PID) controller based on real-time output from the satellite MSS \modif{using the method described in \cite{Chandramohan,CC2019}}
    \item \textbf{C. Machine Learning Controller} – \modif{Uses a pre-trained neural network based on previous MSS to predict the sloshing disturbance and compensate for it in the control feedback loop \cite{fogel2022} \cite{Brunton_Kutz_2022}}.
    \item \textbf{D. Reference Governor} - \modif{A controller that is added onto any pre-existing controller and ensures control inputs and measured outputs remain within safe operational constraints. Input control torques and pressure sensors will be used as the constraints, defining values not to be exceeded \cite{Kolmanovsky-Tutorial}\cite{CFB2022}}.
    \item \textbf{\modif{E. Output Adaptive Feedback or Machine Learning + Reference Governor}} - \modif{Based on the results of experiments B (Output Feedback) and C (ML), select the controller that minimizes the settling time of a given maneuver. In the first case, adaptive control and reference governor are combined using the method of \cite{Magnani}. In the second case, the pre-existing controller uses the pre-trained neural network to compensate for the sloshing disturbances. In doing so, the nominal closed-loop system is linear, and the nonlinear term (i.e the neural network) is transferred to the control input constraint. Thus, the reference governor is designed on a linear system nonlinear constraint \cite{KKG2011}.}
\end{enumerate}

\subsection{Experiment Phases and Concept of Operations}

\begin{figure}[H]
    \centering
    \includegraphics[width=16cm]{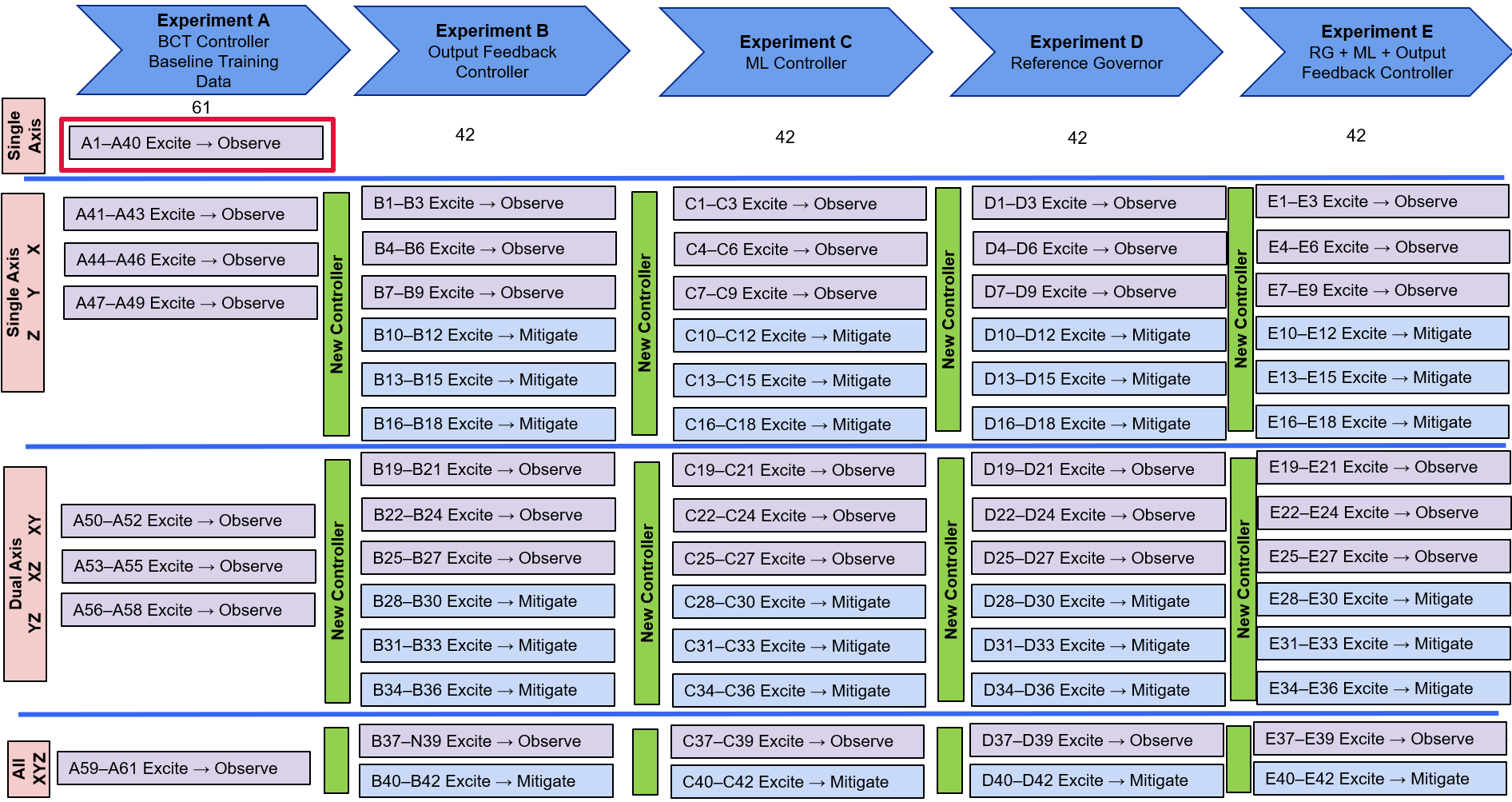}
    \caption{Schematic Diagram Representing Experiment Plan. The red box \modif{indicates} the minimum mission success criteria.}
    \label{fig:experiment_plan}
\end{figure}

The experiment requirements will be carried out across \modif{the five different controllers and each experiment repeated three times, as depicted in Fig. \ref{fig:experiment_plan}. In all,} there will be a total of 229 experiments performed throughout the flight. Experiment Phase A will \modif{include} the minimum success criteria for the entire mission. \modif{Minimum success criteria were determined through simulations which indicated the minimum number of datasets needed to effectively train a neural network model.} 

\begin{figure}[H]
    \centering
    \includegraphics[width=10cm]{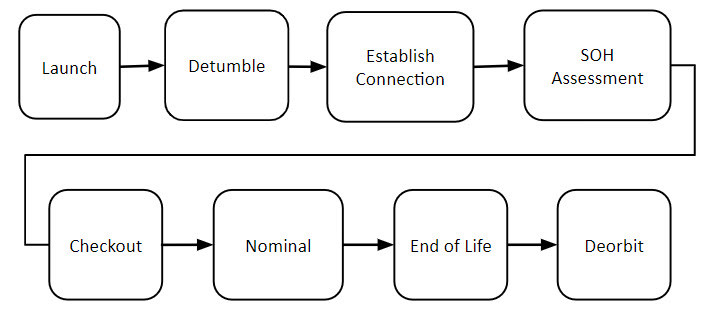}
    \caption{CONOPS SPICEsat lifetime overview.}
    \label{fig:CONOPS}
\end{figure}

\begin{figure}[H]
    \centering
    \includegraphics[width=14cm]{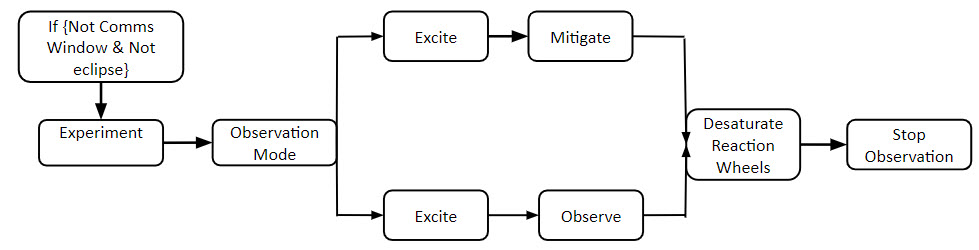}
    \caption{CONOPS Experiment overview block diagram showing SPICEsat operations during each experiment.}
    \label{fig:CONOPS_Exp}
\end{figure}

\begin{figure}[H]
    \centering
    \includegraphics[width=10cm]{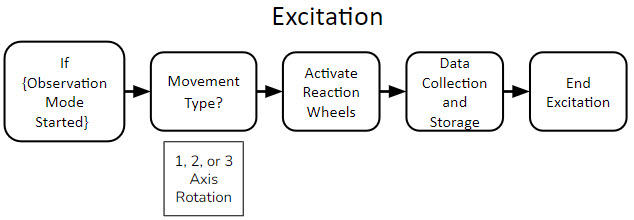}
    \caption{CONOPS Excitation block diagram expanded from Fig \ref{fig:CONOPS_Exp}, showing SPICEsat maneuvering and slosh observations.}
    \label{fig:CONOPS_Excite}
\end{figure}

\begin{figure}[H]
    \centering
    \includegraphics[width=10cm]{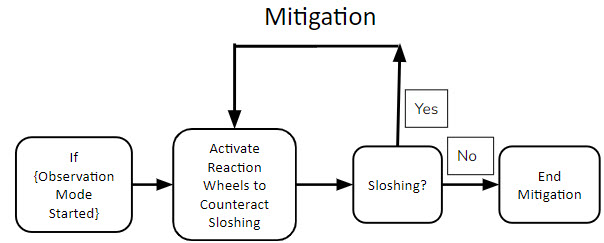}
    \caption{CONOPS Mitigation block diagram expended from Fig \ref{fig:CONOPS_Exp}, showing SPICEsat maneuvering and slosh observations.}
    \label{fig:CONOPS_Mit}
\end{figure}

The SPICEsat mission with 229 experiments expected, will require diligent planning and operations cycles. A detailed  plan has been designed, with the desired maneuvers, software controls, and sensor activations for each experiment. Presented in Figs \ref{fig:CONOPS} - \ref{fig:CONOPS_Mit} are the high level Concept of Operations (CONOPS) for the experimental portions of the mission. Fig \ref{fig:CONOPS} shows the CONOPS block diagram for the mission profile at the highest level, from launch to de-orbit. Figs \ref{fig:CONOPS_Exp}, \ref{fig:CONOPS_Excite}, and \ref{fig:CONOPS_Mit} depict the CONOPS expected for each experiment, and the observation or mitigation branches of CONOPS depending on which experiment is being conducted, respectively, and as indicated in the experiment plan (Fig \ref{fig:experiment_plan}). 

\section{Numerical Simulation Results and Predictive Control Algorithms}

\subsection{Equivalent Mechanical Models}
Attitude changes in a satellite are typically accomplished by the use of reaction control wheels or \modif{magnetorquers}. As a result, we can assume that the mass and the fill ratio of the propellant remains constant. The satellites under consideration are assumed to be in microgravity, so we can neglect the effects of gravity. We also assume that there are no external perturbations, so that $\Gamma_d = 0$. The tank is rigid and in a fixed position. As the satellite changes attitude, the angular velocity and acceleration imparted by the control wheels are the predominant influence on the propellant. 

Following the treatment by Bourdelle \cite{bourdelle2019a} \cite{bourdelle2019b} \cite{bourdelle2019} , we can define an equivalent mechanical model (EMM) to represent the fuel sloshing behavior using a second-order non-linear differential equation:

\begin{equation}
    \ddot{\Gamma}_s(t)  = -A_s(t){\Omega}(t) -B_s(t){\dot\Omega}(t) -C_s(t){\dot\Gamma}_s(t) - K_s(t){\Gamma}_s(t)
    \label{EMM_EoM}
\end{equation}

\begin{equation}
    \dot{\Omega}(t) = \frac{1}{I_{sat}}({\Gamma_{RW}}(t) + {\Gamma_s}(t))
    \label{Omega_dot}
\end{equation}

\begin{figure}[H]
    \centering
	\includegraphics[width=0.45\textwidth]{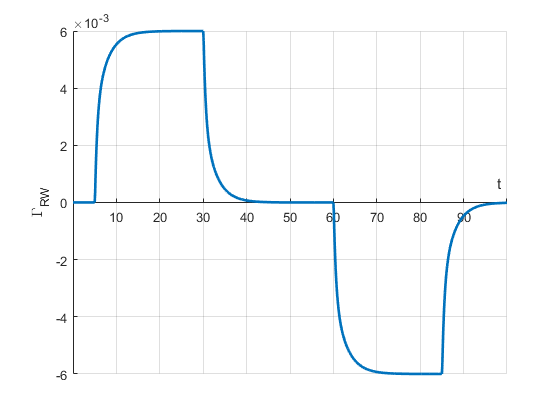}
	\caption{Typical control torque ($\Gamma_{RW}$) with actuator transfer function for a "bang-stop-bang" profile.}
	\label{fig:Tc_RW_typical}
\end{figure}

Input control torques are typically a "bang-stop-bang" profile, where a reaction wheel is spun up and then stopped, imparting its momentum back into the body of the satellite and forcing it to rotate on that axis. Then the reaction wheel is spun up again in the opposite direction until the satellite reaches the desired change of angle and is stopped again, imparting a torque in the opposite direction this time, thereby stopping the satellite rotation. SPICEsat's ADCS is expected to impart up to $\Gamma_{RW} < 0.006$ N-m of torque into the satellite. In practice, these bang-stop-bang profiles are generated by reaction wheels that have a characteristic response curve. We have approximated how these low-pass dynamics are modeled by the a transfer function in Eq. \ref{TransferFunction} taken from \cite{bourdelle2019}. Fig. \ref{fig:Tc_RW_typical} shows the results for a typical input torque profile.

\begin{equation}
    H(s)=\frac{1.2s+0.76}{s^2 + 2.4s + 0.76}
    \label{TransferFunction}
\end{equation}

\begin{figure}[H]
    \centering
	\includegraphics[width=0.6\textwidth]{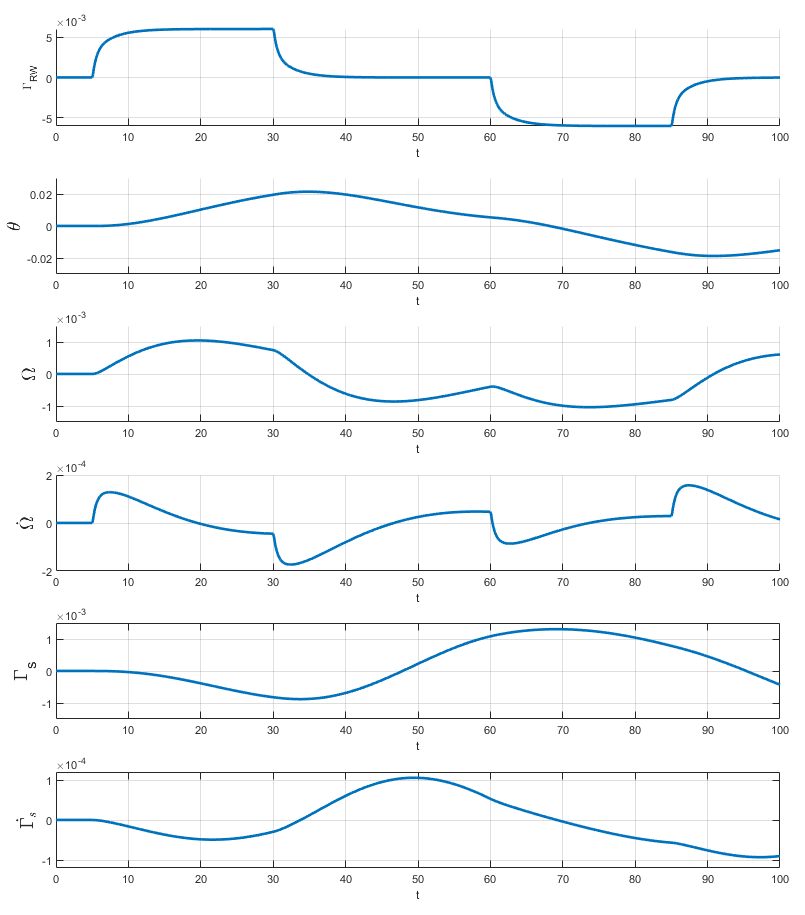}
	\caption{Simulink output showing attitude ($\theta$), angular velocity ($\Omega$), angular acceleration ($\dot\Omega$), sloshing torque ($\Gamma_s$), and sloshing torque dynamics ($\dot\Gamma_s$) for a the typical control torque ($\Gamma_{RW}$) "bang-stop-bang" profile.}
  	\label{fig:slosh_output_typical}
\end{figure}

Following the same methods and parameters as in Bourdelle \cite{bourdelle2019a}, the dynamics from Eq. \ref{EMM_EoM} and \ref{Omega_dot} are modeled in Simulink where we solve for the state vector in Eq. \ref{state_space}. Fig. \ref{fig:slosh_output_typical} represents a typical output from these models. 

\begin{equation}
x(t) \\
= \begin{bmatrix}
        \theta(t)     \\
        \Omega(t)     \\
        \Gamma_s(t)   \\
        \dot\Gamma_s(t) 
    \label{state_space}
\end{bmatrix}
\end{equation}

\subsection{Applying Machine Learning to Satellite Control}
\begin{figure}[H]
    \centering
    \includegraphics[width=0.90\linewidth]{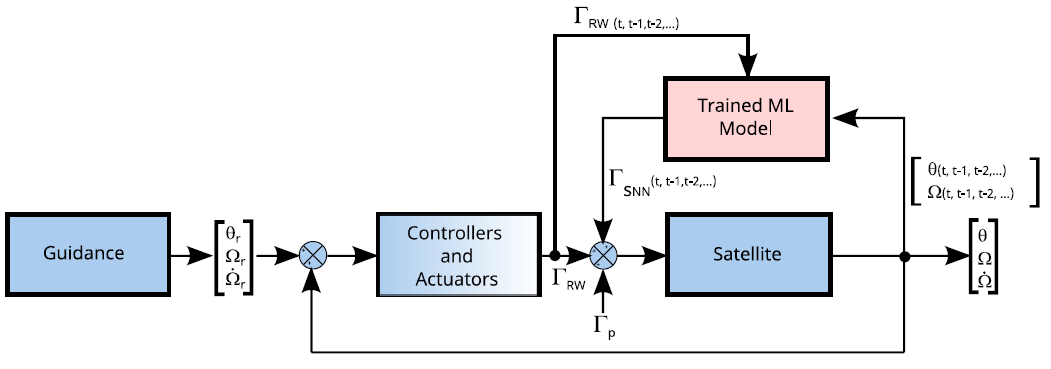}
    \caption{Control block diagram using a ML block for predicting $\Gamma_{s_{NN}}$.}
    \label{fig:control_block_diagram}
\end{figure}

In \cite{fogel2022}, we have examined the feasibility of predicting sloshing behavior onboard satellites during maneuvers using machine learning and using these predictions to design a control algorithm to mitigate the disturbance effect. We do this by training an ML algorithm on a large set of input control torques, $\Gamma_{RW}$, and state vector responses. Once trained, we can replace the EMM in our model with a prediction for the sloshing torque $\Gamma_{s_{NN}}$. We represent this in a simple control block diagram in Fig. \ref{fig:control_block_diagram}.

In this case, the dynamics of the system are completely replaced by a trained "black box" knowing nothing about the physics of the problem. The ML model is trained offline from the control algorithm. This "Black Box" takes inputs from the system ($\theta, \Omega, \Gamma_{RW}$) and predicts the effect of sloshing on the system, such that $\Gamma_{s_{NN}} \approx \Gamma_s$. This sloshing prediction is then \modif{fed back into}  the existing PID controller, and a new control torque $\Gamma_{RW}$ is calculated.

\subsection{\modif{Designing the Machine Learning Model}}

Machine Learning requires a set of training data to learn the dynamics of the problem. We can excite several different sloshing modes $\Gamma_s(t)$ by creating different control torque values, $\Gamma_{RW}(t)$, running several models using Matlab Simulink, and stacking these results as input to the ML model. This generates a large data set of control inputs and output state vectors $x = [ \theta, \Omega, \Gamma_s \dot{\Gamma_s}] $ to train against. Next, we are left with the selection of exactly what  $\Gamma_{RW}$ values to \modif{model}. Recalling the bang-stop-bang, we can consider:

\begin{enumerate}
    \item Strength (torque) of the reaction wheel torque ($\Gamma_{RW}$)
    \item Duration of the input control torque "bang" ($t_{dur}$)
    \item Dwell time between the first "bang" and the equal and opposite second "bang" ($t_{dwell}$)
\end{enumerate}

The strength of the input control torque is governed by the physical characteristics of the reaction wheels. In the case of the Blue Canyon Technologies (BCT) XACT-50, $\Gamma_{RW}(max)$ = 0.006 N-m.  With knowledge of the physical characteristics and taking similar values as Bourdelle \cite{bourdelle2019b}, we choose the following parameters, which result in typical torque profiles in Fig. \ref{fig:Tc_RW_typical}.

\begin{equation}
    0.002 \text{ N-m} < \Gamma_{RW} < 0.006 \text{ N-m}
    \label{Trw_values}
\end{equation}

\begin{equation}
    25 \text{ s} < t_{dwell} < 40 \text{ s}
    \label{tdwell}
\end{equation}

\begin{equation}
    5 \text{ s} < t_{dur} < 25 \text{ s}
    \label{tdur}
\end{equation}

\subsection{\modif{Machine Learning Algorithm Selection}}
\modif{
In the broad and growing field of ML, there are a large number of ML algorithms that can be applied. We chose five different neural network algorithms to investigate. In the Supplemental Materials, we provide details used in each ML algorithm proposed here:
\\
\vspace{-0.25cm}
\begin{enumerate}
    \item Sigmoid (SIG) \cite{NARAYAN199769}
    \item Radial Basis Function Neural Networks (RBFNN) \cite{chen1991orthogonal}
    \item Regression Trees (RT) \cite{breiman2001}
    \item Artificial Neural Networks (ANN) \footnote{Generate Feed Forward Neural Network, The MathWorks Inc., MATLAB version: R2022a, \url{https://www.mathworks.com/help/deeplearning/ref/feedforwardnet.html}, accessed 06-30-2024}
    \item Nonlinear Autoregressive Exogenous Model (NARX) \cite{Billings}
\end{enumerate}}

\subsection{\modif{Machine Learning Results}}

\begin{figure}[H]
    \centering
    \includegraphics[width=0.61\linewidth]{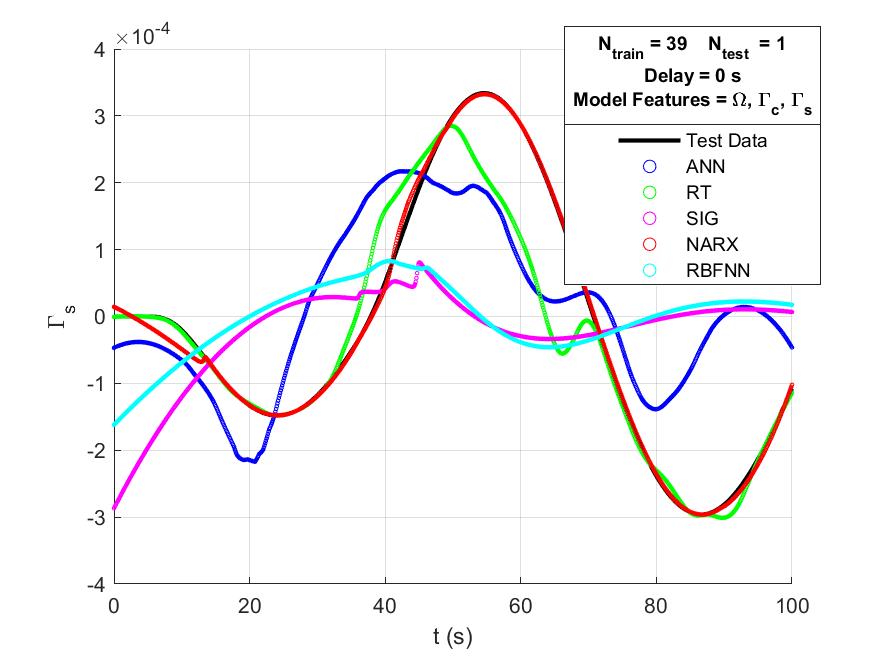}
    \caption{Different neural network fit predictions to a single test data set for no input delay case. \modif{The black line is the desired curve to fit.}}
    \label{fig:ML_predictions}
\end{figure}

After the state space data is loaded into the associated data sets and the training algorithms designed, the time-consuming training process against the data begins \modif{as detailed in the Supplemental Materials provided with this paper}. Fig. \ref{fig:ML_predictions} shows a typical result of fitting the 5 different algorithms to two of the measured states, $[\Omega, \Gamma_{RW}]$ and attempting to predict the slosh Response, $\Gamma_{s{NN}}$. The black line in Fig. \ref{fig:ML_predictions} represents the goal the NN models are trying to fit. 

These numerical results show that NARX neural networks (Nonlinear Autoregressive Exogenous Models) are best suited for application to this problem. This makes sense because NARX networks are commonly used in time-series modeling. For SPICEsat, we propose to use the actual excitation data recorded onboard the satellite. \modif{SPICEsat angular acceleration and angular velocity data along with the associated control inputs will be used as training inputs to the NARX neural network. The resulting NARX model will output the predicted sloshing torque, $\Gamma_{s_{NN}}$. This $\Gamma_{s_{NN}}$ will be feedback into the control torque to compensate for the slosh disturbance. A typical PID control law to command SPICEsat to a desired angle, $\theta_d$, with sloshing mitigation would become:}   

\begin{equation}
    \Gamma_{RW}= -k_p(\theta - \theta_d) - k_d \Omega - \Gamma_{s_{NN}}
    \label{Trw_with_Tnn_correction}
\end{equation}

\modif{In Eq. \ref{Trw_with_Tnn_correction}, $k_p$ and $k_d$ are tunable gains that can be determined using classical linear control methods (for example, pole placement) assuming the sloshing disturbance has been compensated.}

\subsection{\modif{Key Machine Learning Assumptions and Limitations}}

\begin{figure}[H]
    \centering
    \includegraphics[width=0.70\linewidth]{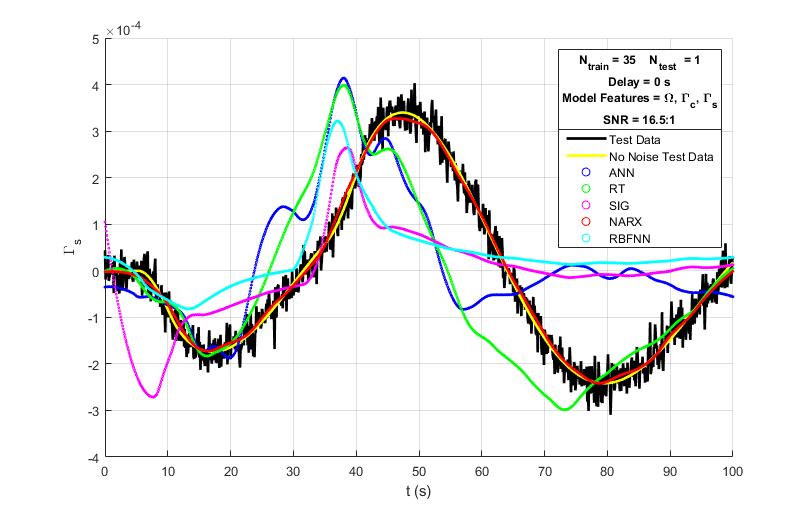}
    \caption{\modif{Neural network slosh predictions in the presence of a noisy signal (SNR=16.5:1). The black line is the desired curve to fit.}}
    \label{fig:ML_predictions_noise}
\end{figure}

\modif{In Fogel \cite{fogel2022}, all Machine Learning algorithms were tested the NARX algorithm was tested for robustness using various signal-to-noise ratio (SNR) cases. Specifically the angular velocity, angular acceleration, and sloshing torque measurements were treated with SNR ratios as low as 16.5:1 \cite{MEMS_sensors_noise} as shown in Fig. \ref{fig:ML_predictions_noise}. This analysis shows that with smaller SNR ratios, each prediction performs successively worse, however, the NARX model holds up considerably well in the presence of excessive noise. Noisy data does result in an element of overfitting (ripples) in the NARX model. Overfitting is caused by the neural network trying to learn "too much" from the now noisy input data. In our research, the traditional feed-forward neural network (ANN), radial basis function (RBFNN), and sigmoid fitting (SIG) networks perform poorly in the presence of noise. Regression trees perform somewhere between the NARX models and ANNs, although the fit seems to lag the test data. A second limitation of the NARX model indicates that the prediction of sloshing results is best when the control inputs and angular rates are within the ranges that the model was initially trained. A simple way to look at this is that the NARX model is very effective in interpolation but not extrapolation outside the regime on which it was trained.}

\section{Computational Fluid Dynamics Modeling of SPICEsat}
\subsection{Satellite Design Overview}
\begin{figure}[H]
    \centering
    \includegraphics[width=16.5cm]{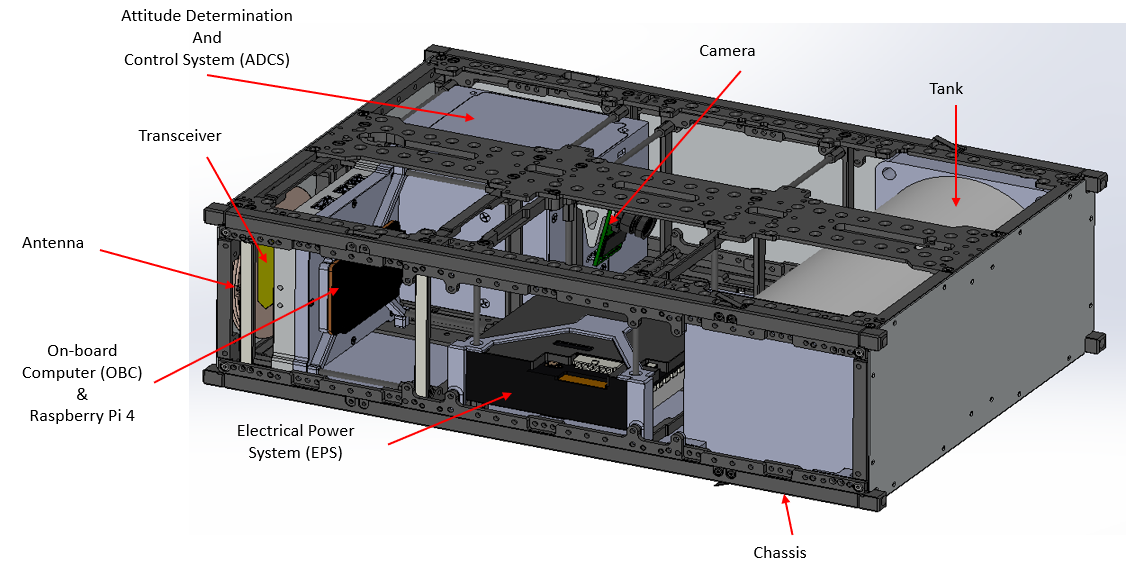}
    \caption{Satellite Preliminary Design Concept}
    \label{fig:sat_design}
\end{figure}

CFD modeling of SPICEsat requires a number of inputs to be defined and configured. We use FLOW-3D from FlowScience as the primary tool for this research. In Section II.C, we define the main physical characteristics of the fluid \modif{and the} tank. To date, SPICEsat has been through a number of design phases focusing on the individual components required to fulfill the mission. Fig. \ref{fig:sat_design} and Table \ref{table:components} show the most recent satellite configuration with key components included. Solar panels are removed in order to show a more complete internal diagram.  Each sensor component has been analyzed and compared against the mission requirements discussed in Section II. \modif{Trade studies against these requirements are presented in Section IV.E.
\\
SPICEsat is designed as a platform for control software experimentation. This fact had significant ramifications for the selection of an onboard computer (OBC). The OBC would be required to handle real-time sensor data processing, calculations, and output of satellite control commands for each experiment. Most commercial OBCs run at lower process speeds and therefore have less capability when it comes to numerical computation. In consideration this fact and in consultation with small satellite industry experts, the authors decided to separate processors into an Onboard Computer for satellite management and a Payload Computer to conduct the experiment. A GOMSpace NanoMind A3200 was selected as the OBC and a Rasberry Pi 4B+ was chosen as the Payload Computer. Table \ref{table:obc_components} compares the two different computer selections.}

\begin{table}[H]
  \centering 
  \caption{SPICEsat Component Overview} 
  \begin{tabular}{l l l}
    \hline\hline 
    \textbf{Component} & \textbf{Description} & \textbf{Vendor}  \\ 
    \hline 
    ADCS (MSS)              & Satellite attitude and control & Blue Canyon XACT-50\\
    Camera (VSS)            & Sloshing video recording       & ArduCam OV5647  \\
    Tank                    & Primary experiment with fluid  & Polycarbonate \\
    Pressure Sensors (LSS)  & Pressure sensor mapping        & Sensitronics \\
    Chassis \& Panel        & Satellite main chassis         & SpaceMind 6U \\
    Electrical Power        & Power for satellite            & SpaceMind \\
    Payload Computer        & Payload computer / control     & RasberryPi 4B+ \\
    Onboard Computer        & Flight computer                & GOMSpace NanoMind A3200 \\
    UHF Transceiver         & Omni-directional comms         & Ultra High Frequency (UHF) Transceiver \\
    S-band Transceiver      & High data rate comms           & GOMSpace Software Defined Radio (SDR) \\
    \hline 
  \end{tabular}
  \label{table:components}
\end{table}

\begin{table}[H]
  \centering 
  \begin{threeparttable}
  \caption{\modif{SPICEsat Computer Specifications}} 
  \label{table:obc_components}
  \begin{tabular}{l l l l l l}
    \hline\hline 
    \textbf{Component} & \textbf{Description} & \textbf{Processor} & \textbf{Speed} & \textbf{Memory} & \textbf{Interfaces}\\ 
    \hline 
    Onboard Computer        & GOMSpace A3200          & AVR32                & 64 MHz       &  32MB & I2C \tnote{1}, UART \tnote{2}, CAN\tnote{3}, GPIO\tnote{4}\\
    Payload Computer        & RasberryPi 4B+          & Quad Cortex-A72 & 1.8 GHz           & 8 GB  & I2C, Serial, GPIO \\
     \hline 
\end{tabular}
\begin{tablenotes}
    \item [1] I2C is a synchronous, multi-controller/multi-target, single-ended, serial communication bus
    \item [2] UART is a Universal Asynchronous Receiver-Transmitter (UART) communications bus
    \item [3] CAN is a controller area network (CAN) communications bus
    \item [4] GPIO refers to general purpose input / output (GPIO) pins typical on a RaspberryPi
\end{tablenotes}
\end{threeparttable}
\end{table}

\subsection{Input Torque Modeling}
The amount of available torque we can impart into the satellite is determined by the reaction wheel momentum delivered by the ADCS. The choice of the Blue Canyon Technologies XACT-50 is the starting point for the CFD modeling. The product data sheet \cite{BCT_data_sheet} indicates the maximum reaction wheel torque output by the device is $\Gamma_{RW} = 0.006$ Nm and up to 10 deg/sec slew rate for a 14kg, 6U CubeSat. Notable here is that SPICEsat is modeled at between 7 - 8 kg, meaning with the given input torque, we expect a higher spin rate, which we \modif{have} modeled at 16 deg/sec. This assumption has been confirmed in discussions with Blue Canyon Technologies, \modif{who project that the experiment could get rates as high as 26 deg/sec}. 

\begin{figure}[H]
    \centering
    \includegraphics[width=10cm]{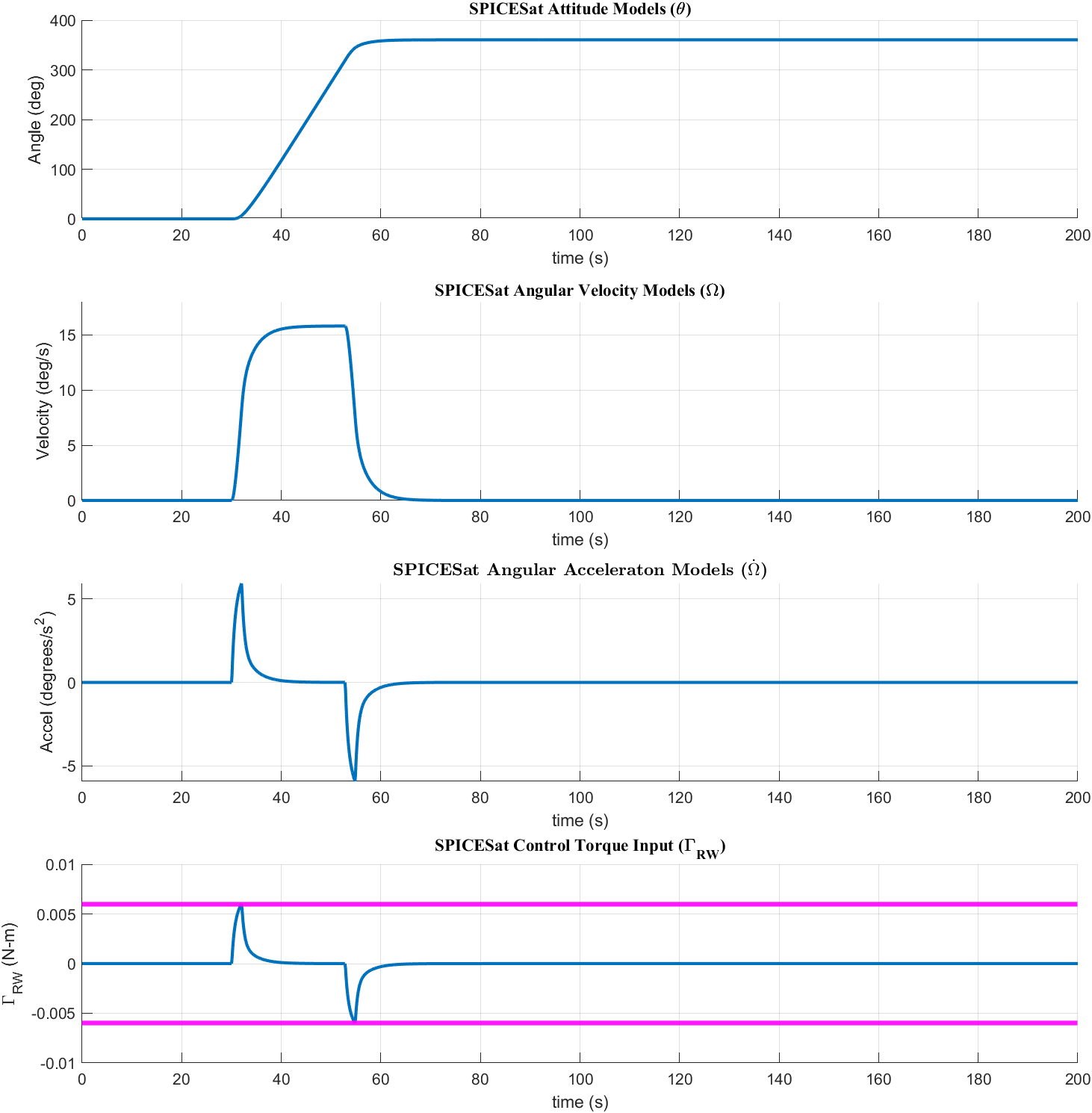}
    \caption{Reaction Wheel Torque, $\Gamma_{RW}$ used in the CFD models.}
    \label{fig:rw_torque}
\end{figure}

The input control is a reaction wheel modeled by the transfer function taken from \cite{bourdelle2019} and is given in Eq. \ref{TransferFunction}. Assuming a satellite moment of inertia (MoI) of $I_{sat} = 0.058$ kg-m$^2$, we command the satellite to rotate by 360 \degree using a maximum torque of $\Gamma_{RW} (max) = 0.006$ N-m. Then we can calculate the resulting angular acceleration profile $\dot{\Omega}(t)$ using Eq. \ref{IsatOMegat-Trw}. Fig. \ref{fig:rw_torque} shows a typical attitude control profile used in the CFD modeling. 

\begin{equation}
    I_{sat}\dot{\Omega}(t) = \Gamma_{RW}(t)
    \label{IsatOMegat-Trw}
\end{equation}

As depicted in Fig. \ref{fig:MOI-Coordinates}, we can impart this angular acceleration profile, $\dot{\Omega}(t)$, \modif{along} one of three different axes, $X_{BODY}$, $Y_{BODY}$, or $Z_{BODY}$, \modif{or along} two or three axes at once. In Section IV.D, we present the results of each axis independently and of all three axes at once (i.e., a tumbling motion). 

\subsection{CFD Modeling}
For our study, we chose to use Flow-3D from FlowScience. Flow-3D is a computational fluid dynamics solver capable of mimicking flow in zero-g. The analysis begins with importing the satellite tank CAD model into the software, defining a computational mesh within the software, defining the fluid we want to model, and establishing initial conditions for the fluid. In Table \ref{table:CFD Parameters}, we define the key parameters of our study.

\begin{table}[ht]
  \centering 
  \caption{CFD Analysis Parameters} 
  \begin{tabular}{c c l}
    \hline\hline 
    \textbf{Parameter} & \textbf{Measure} & \textbf{Description}  \\ 
    \hline 
    Fluid               & 70\% Full   & Water at 20\degree C  \\
    Gravity             & Zero-g      & Zero-gravity in space\\
    $\Gamma_{RW}$       & rad/s       & Commanded control torque (See Sec IV.B) \\
    Rotation Center     & 0.15 m      & Distance between satellite center of mass and center of tank \\
    Fluid Contact Angle & 40 \degree  & Measure of wetting of a solid by a liquid \\
    Mesh                & 0.003 m     & Computational mesh grid resolution \\
    Mesh                & 99,937 pts  & Computational mesh grid size \\
    \hline 
  \end{tabular}
  \label{table:CFD Parameters}
\end{table}

\begin{figure}[H]
    \centering
    \includegraphics[width=10cm]{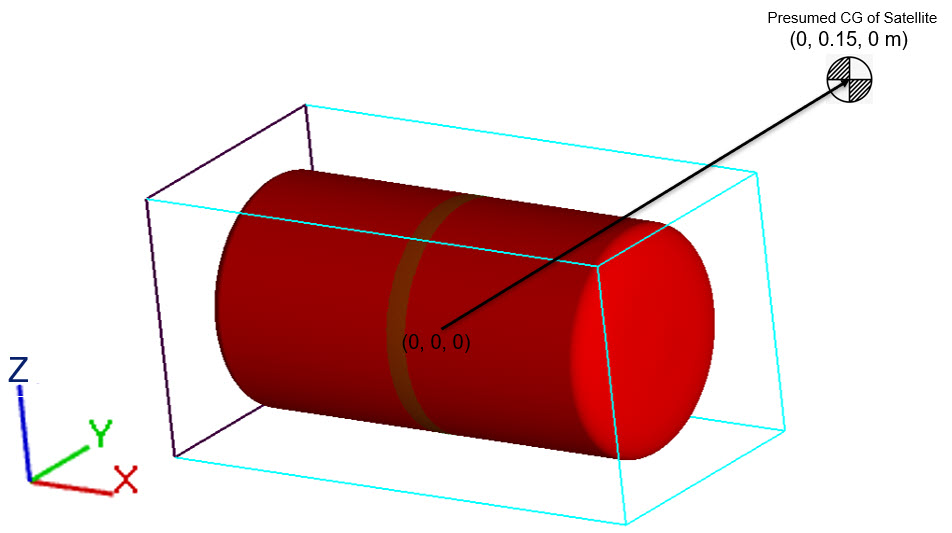}
    \caption{SPICEsat / Satellite Body Coordinate Systems.}
    \label{fig:SPICEsat_Coord}
\end{figure}

Next, we consider initial conditions. The Florida Institute of Technology SPHERES experiment onboard the ISS discovered that the complex initial conditions consisting of large numbers floating bubbles, non-uniformly distributed throughout the tank, are difficult to reproduce accurately in CFD \cite{Florida_SPHERES}. Their experiment considered three different initial maneuvers to set the fluid in a repeatable set of initial conditions, finally settling on a simple maneuver of spinning the tank along a single axis, leaving the liquid positioned equally at both ends of the tank. Since the tank is offset from the center of mass (CoM) of the satellite coordinate system, we place the center of the tank at 0.15 m from the CoM of the satellite itself, as shown in Fig \ref{fig:SPICEsat_Coord}. The Flow3D is then rotated about the satellite CoM. 

\subsection{CFD Results}
\begin{figure}[H]
\centering
\begin{tabular}{r p{\subfigwidth} p{\subfigwidth} p{\subfigwidth}}

\textbf{$\Gamma_{RW}$ (X axis)} & 
\begin{subfigure}[b]{\subfigwidth}
    \caption{t = 25 s}
    \includegraphics[width=\subfigwidth]{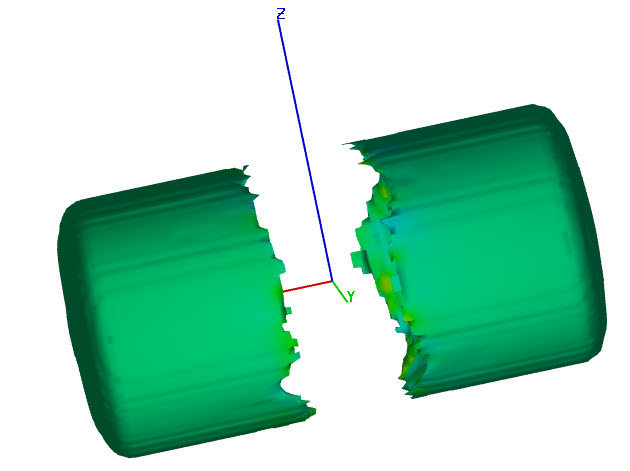}
\end{subfigure} &
\begin{subfigure}[b]{\subfigwidth}
    \caption{t = 36 s}
    \includegraphics[width=\subfigwidth]{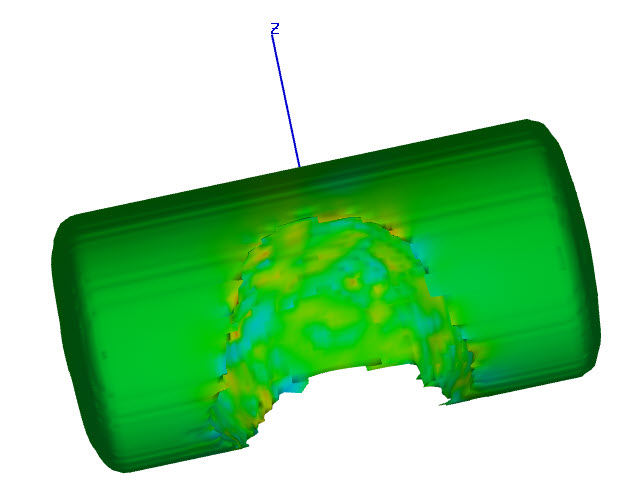}
\end{subfigure} &
\begin{subfigure}[b]{\subfigwidth}
    \caption{t=100s}
    \includegraphics[width=\subfigwidth]{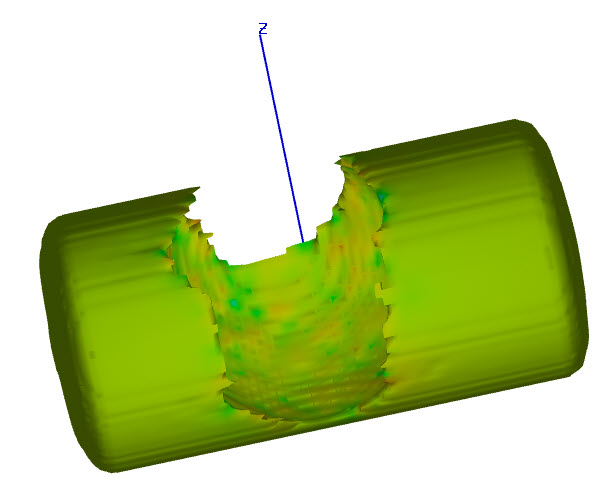}
\end{subfigure} \\

\textbf{$\Gamma_{RW}$ (Y axis)} & 
\includegraphics[width=\subfigwidth]{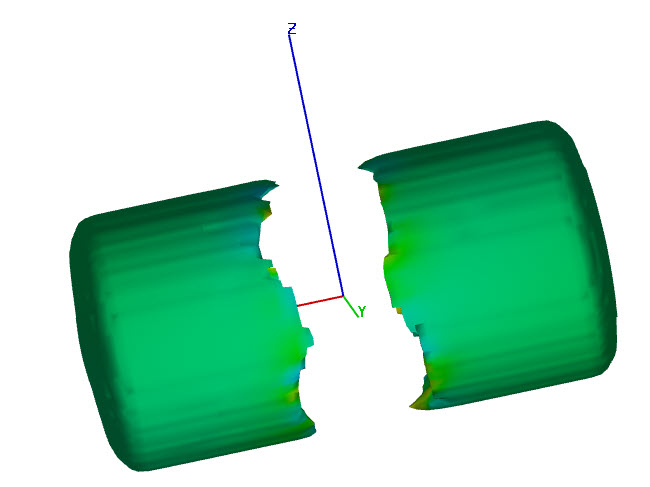} &
\includegraphics[width=\subfigwidth]{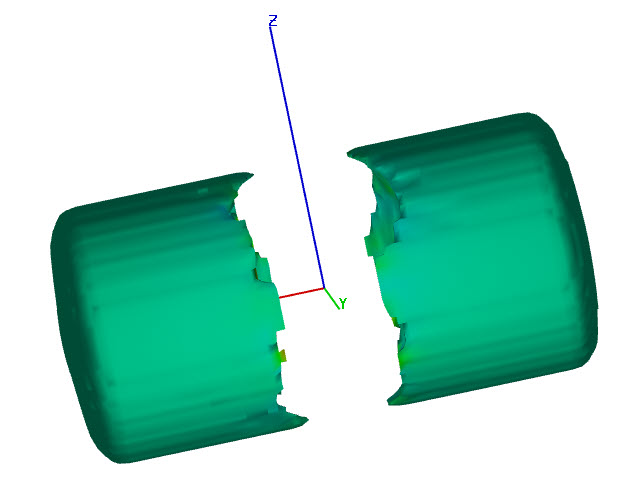} &
\includegraphics[width=\subfigwidth]{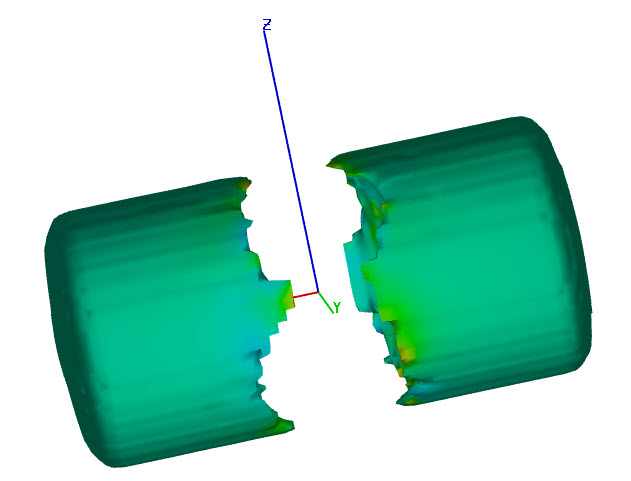} \\

\textbf{$\Gamma_{RW}$ (Z axis)} & 
\includegraphics[width=\subfigwidth]{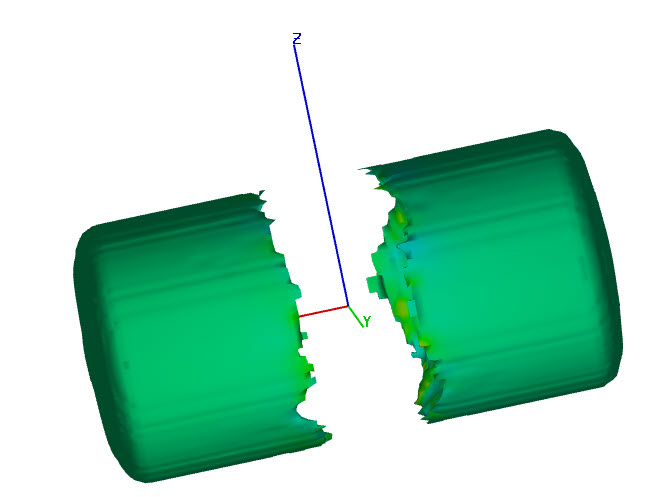} &
\includegraphics[width=\subfigwidth]{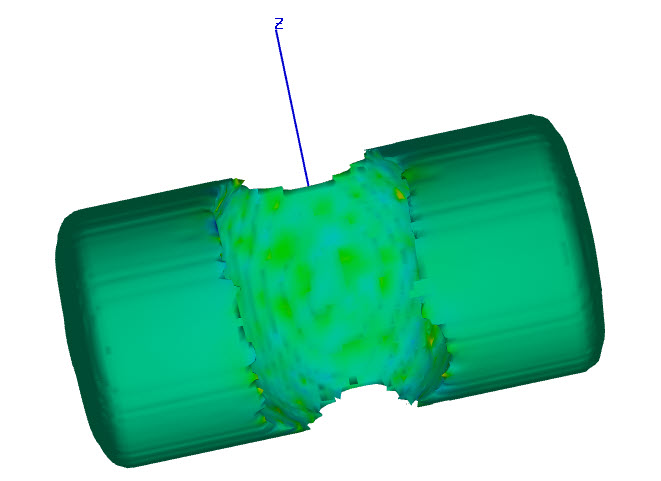} &
\includegraphics[width=\subfigwidth]{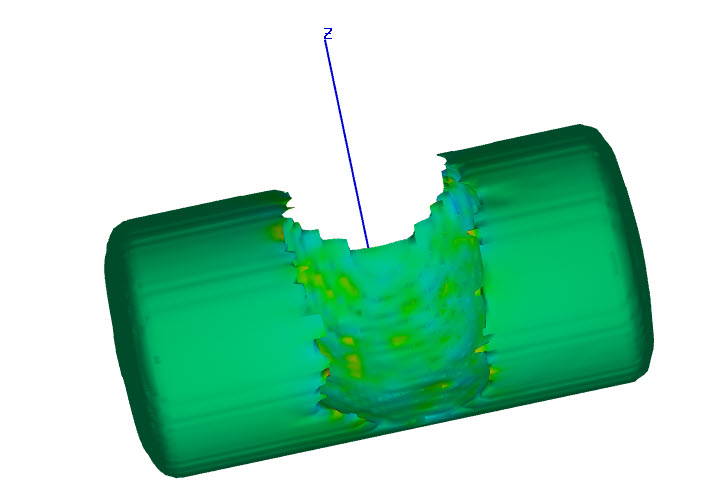} \\

\textbf{$\Gamma_{RW}$ (XYZ axes)} & 
\includegraphics[width=\subfigwidth]{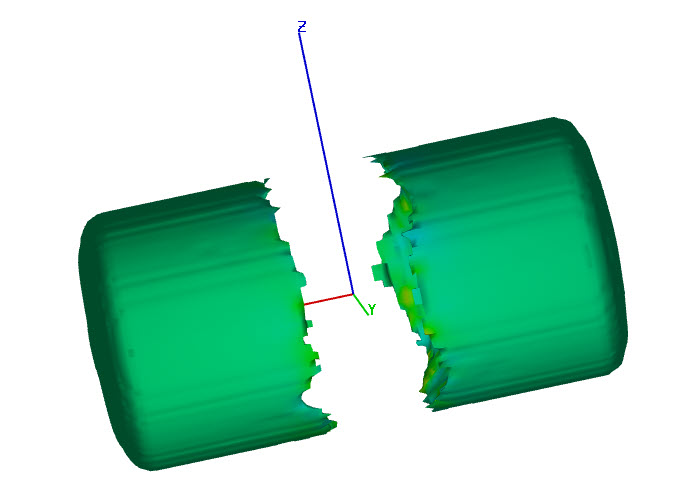} &
\includegraphics[width=\subfigwidth]{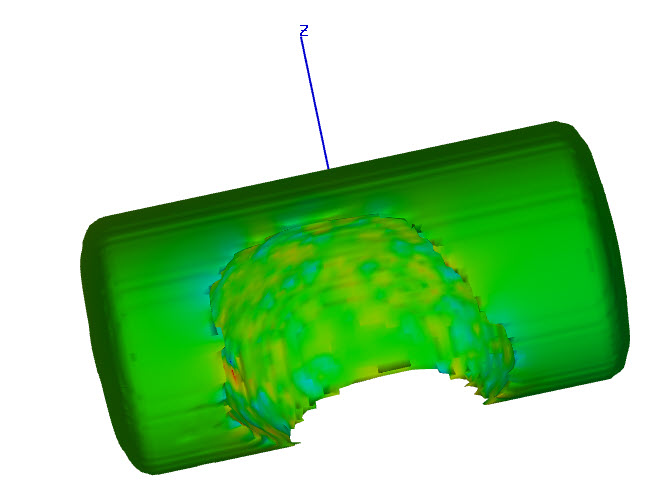} &
\includegraphics[width=\subfigwidth]{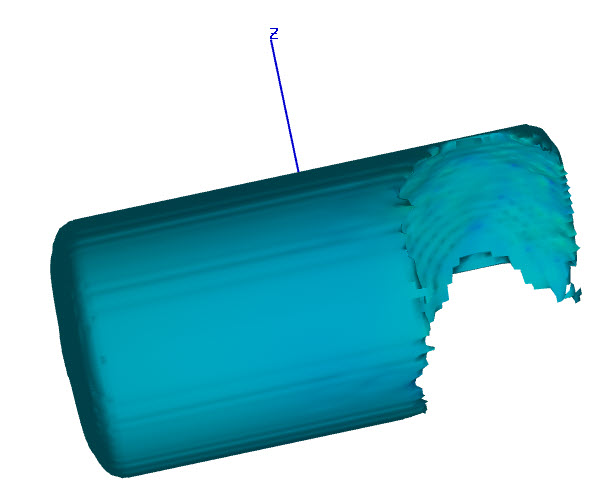} \\

\end{tabular}

\caption{CFD Results at t = 25s (initial state), t = 36s (Max $\Gamma_{RW}$), and t = 100s for input torques along the X,Y,Z, and XYZ axes.}
    \label{fig:CFD_Results}
\end{figure}

With all the preliminaries established, four different control torques are input into the CFD simulations to model the Blue Canyon Technologies maximum torque around the \textit{x,y,z}, and simultaneous \textit{xyz} axes (i.e., tumble maneuver). Fig. \ref{fig:CFD_Results} shows the results of the simulations. As indicated in Fig. \ref{fig:rw_torque}, no input torque is given until t = 30 s, so the images in Fig. \ref{fig:CFD_Results} all show the initial conditions at t = 25 s with the fluid at rest. The maximum torque (bang) of $\Gamma_{RW}$(max) = 0.006 N-m is exerted at t = 36 s. At t = 46 s, an opposite input is exerted ($\Gamma_{RW}$(max) = - 0.006 N-m), and finally the fluid position at t = 100 s. Significant sloshing is detected when the torque is input around the \textit{x} and \textit{z} axes, but very little disturbance is noted about the \textit{y} axis. This is to be expected, as the initial condition is only reinforced by a rotation about the \textit{y} axis. The most significant disturbance occurs in the simulated tumbling motion case. These simulations also indicate that the fluid still has not settled at t = 100 s.

\begin{figure}[H]
\centering
\begin{tabular}{r p{\subfigwidthb} p{\subfigwidthb} }

\rotatebox[origin=l]{90}{\textbf{A) $\Gamma_{RW}$ Along X axis}}
\begin{subfigure}[b]{\subfigwidthb}
    \caption{Sloshing Forces (N)}
    \includegraphics[width=\subfigwidthb]{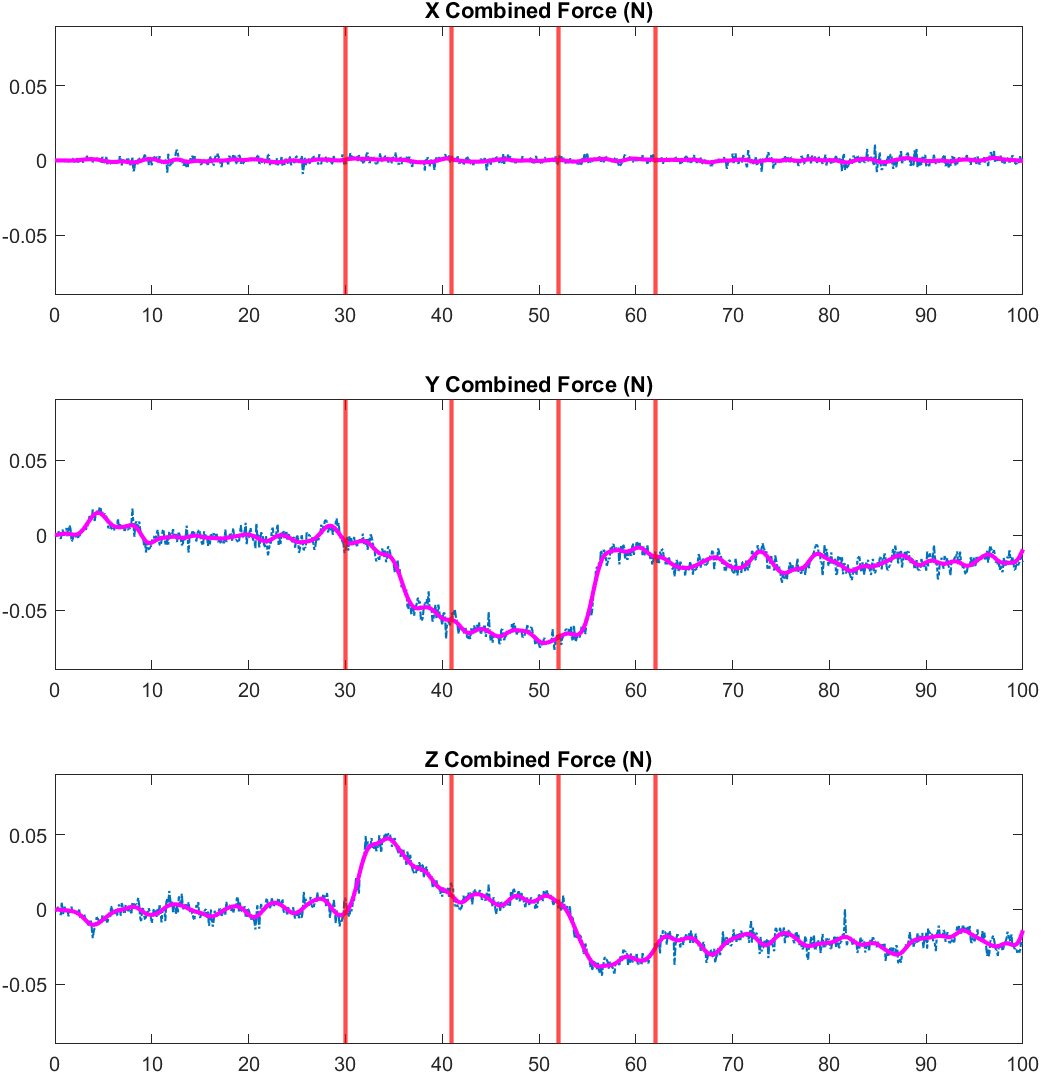}
\end{subfigure} &
\begin{subfigure}[b]{\subfigwidthb}
    \caption{Sloshing Torque $\Gamma_s$ (N-m)}
    \includegraphics[width=\subfigwidthb]{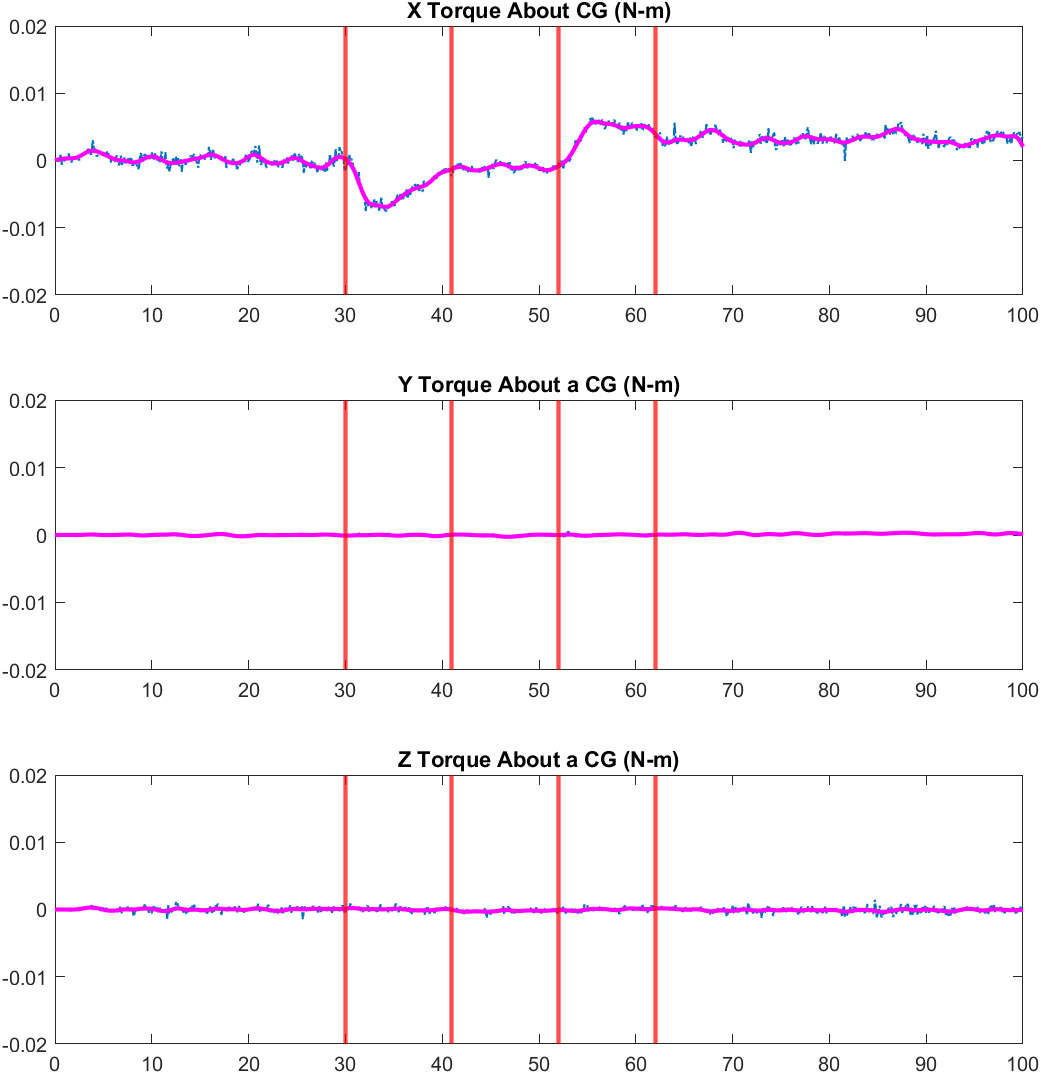}
\end{subfigure} \\

\rotatebox[origin=l]{90}{\textbf{B) $\Gamma_{RW}$ Along Y axis}}
\includegraphics[width=\subfigwidthb]{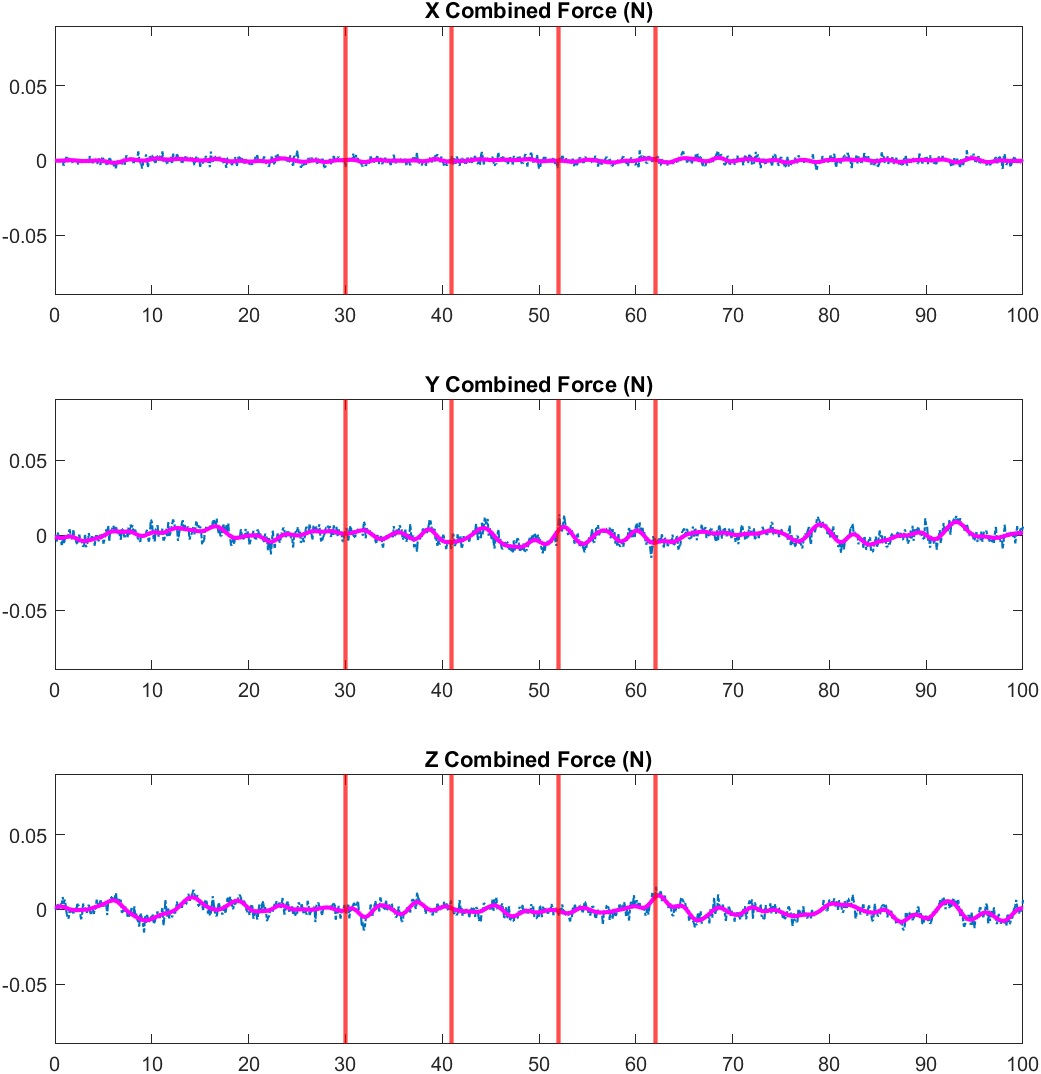} &
\includegraphics[width=\subfigwidthb]{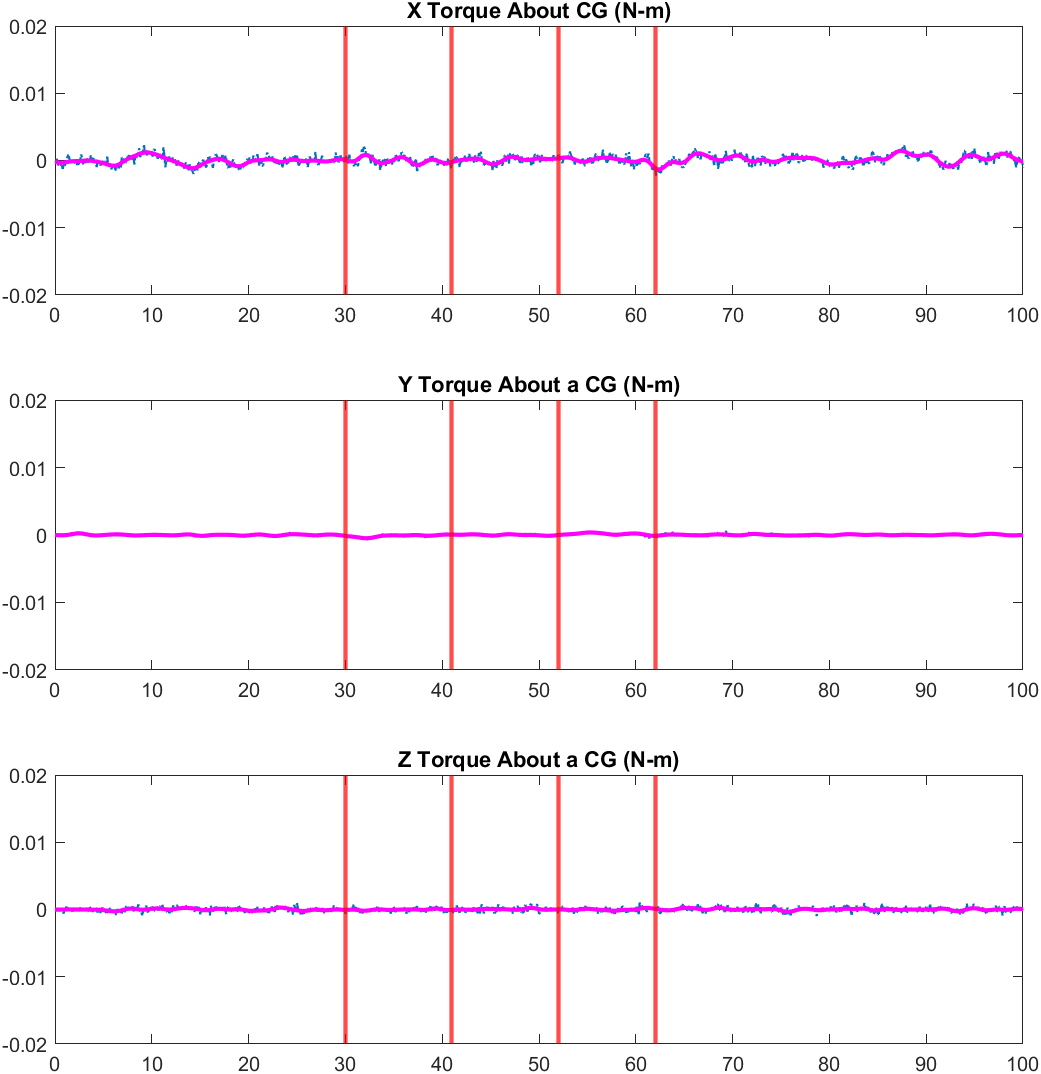} \\

\end{tabular}

\caption{Sloshing disturbance Forces and Torques along all three axes.  A), B) represent input torques,($\Gamma_{RW}$), along the $X$ and $Y$ axes respectively. Vertical lines indicate where the input control torques begin/end. The blue dotted line is the raw CFD output. The magenta line is a low pass filtered smoothing function.}
\label{fig:Forces_Torques1}
\end{figure}

\begin{figure}[H]
\centering
\begin{tabular}{r p{\subfigwidthb} p{\subfigwidthb} }

\rotatebox[origin=l]{90}{\textbf{C) $\Gamma_{RW}$ Along Z axis}}
\begin{subfigure}[b]{\subfigwidthb}
    \caption{Sloshing Forces (N)}
    \includegraphics[width=\subfigwidthb]{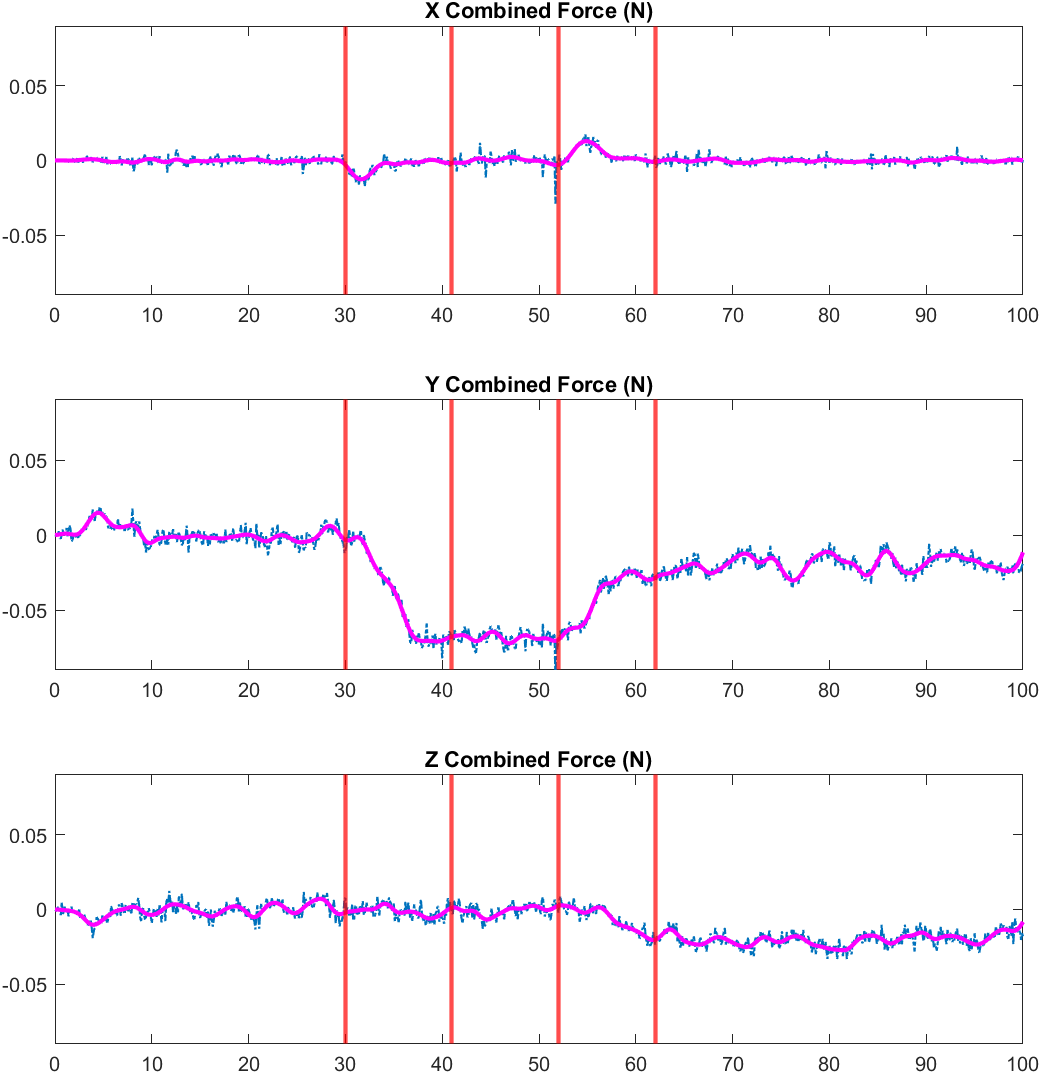}
\end{subfigure} &
\begin{subfigure}[b]{\subfigwidthb}
    \caption{Sloshing Torque $\Gamma_s$ (N-m)}
    \includegraphics[width=\subfigwidthb]{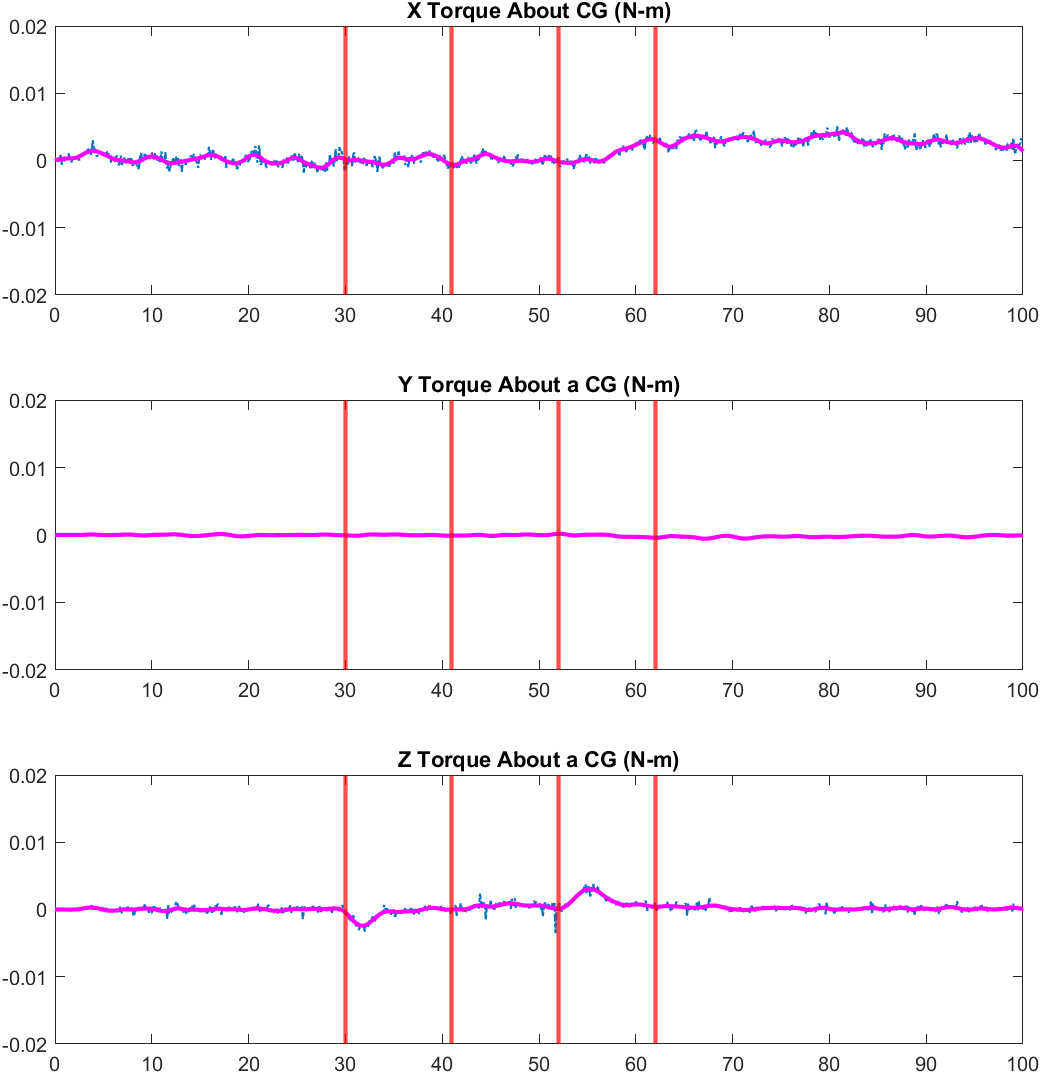}
\end{subfigure} \\

\rotatebox[origin=l]{90}{\textbf{D) $\Gamma_{RW}$ Along XYZ axis}}
\includegraphics[width=\subfigwidthb]{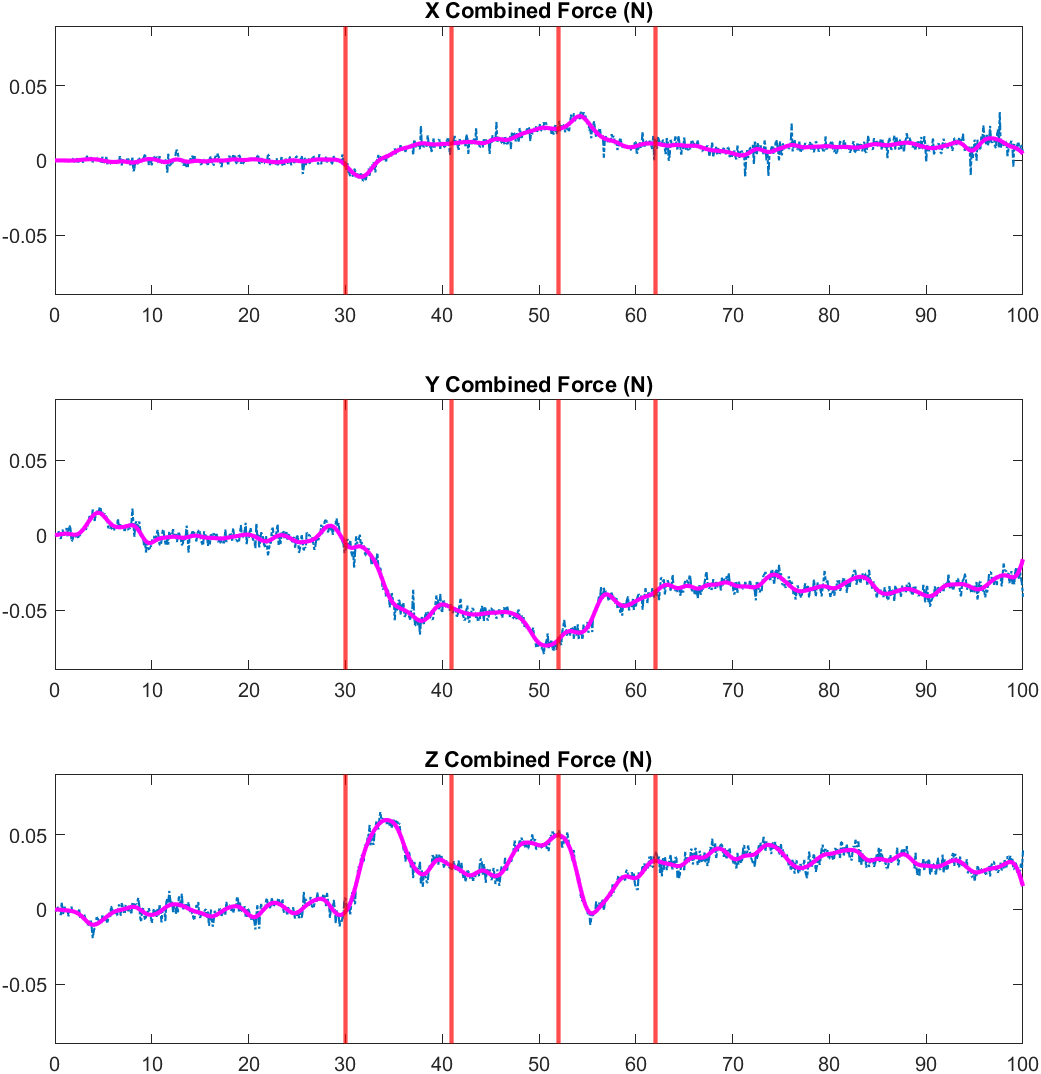} &
\includegraphics[width=\subfigwidthb]{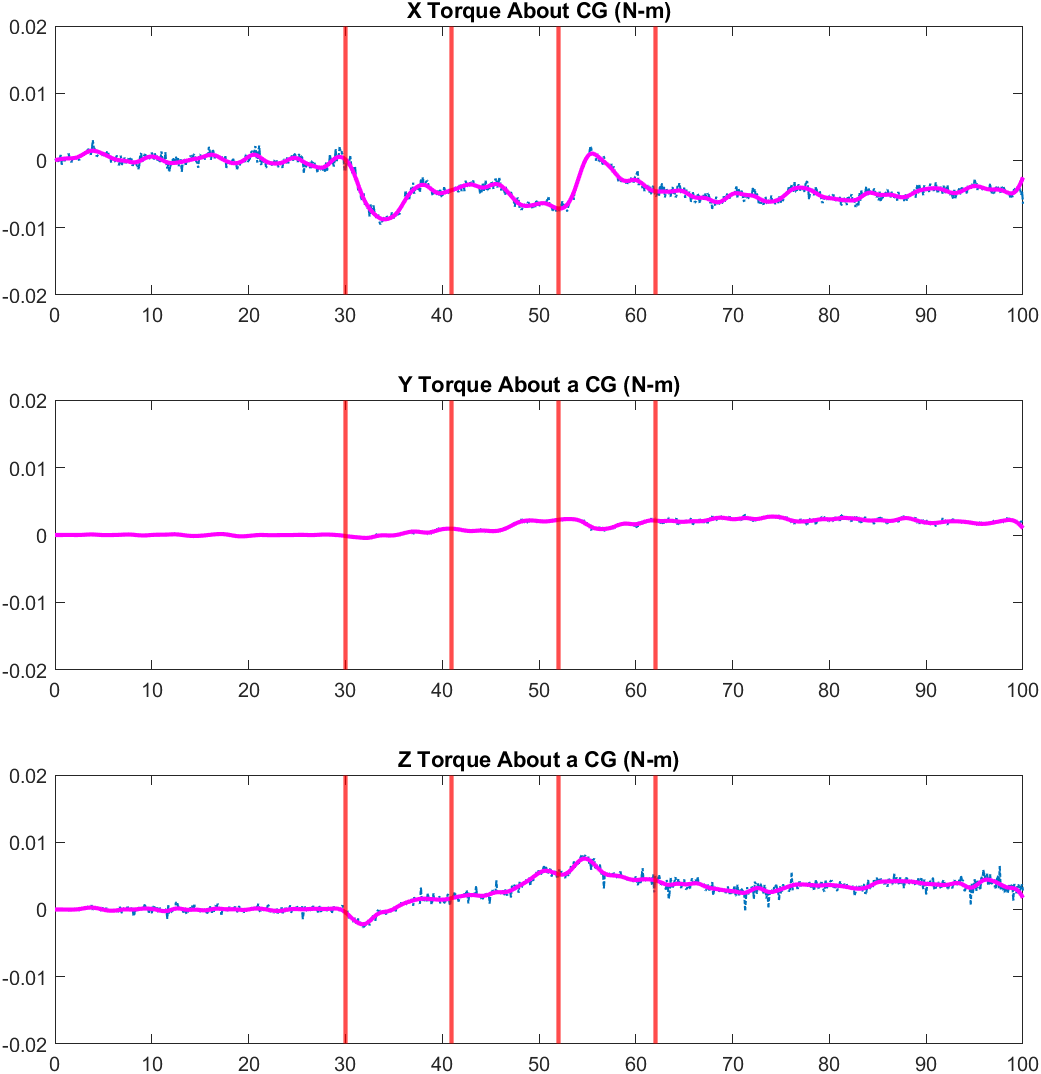} \\

\end{tabular}

\caption{Sloshing disturbance Forces and Torques along all three axes.  C), D) represent input torques,($\Gamma_{RW}$), along the $Z$ and $XYZ$ axes respectively. Vertical lines indicate where the input control torques begin/end. The blue dotted line is the raw CFD output. The magenta line is a low pass filtered smoothing function.}
\label{fig:Forces_Torques2}
\end{figure}

The CFD analysis calculates the resulting forces and torques in each direction that are generated by the sloshing disturbance in the tank. We present those here in Fig. \ref{fig:Forces_Torques1} and Fig. \ref{fig:Forces_Torques2}.

Since we desire our satellite gyroscopes and pressure sensors detect the sloshing, we examine data from Fig. \ref{fig:Forces_Torques1} and \ref{fig:Forces_Torques2} to determine the maximum sloshing disturbance in all four cases and summarize this data in Table \ref{table:max_disturbances}.

\begin{table}[H]
\caption{\label{table:max_disturbances} Sloshing Disturbance Forces and Torques Derived from CFD Calculations}
\centering
\begin{tabular}{ccc}
\hline
$\Gamma_{RW}$ Direction & Max F(N)&  Max $\Gamma_{s}$(N-m)\\ \hline
X-Axis  & 4.75E-2  & 5.77E-3\\
Y-Axis  & 9.35E-3  & 1.44E-3\\
Z-Axis  & 1.48E-2  & 4.17E-3\\
XYZ-Axis& 5.99E-2  & 7.56E-3\\
\hline
\end{tabular}
\end{table}

\subsection{MSS, LSS, and VSS Design Requirements and Trade Studies}

\subsubsection{Motion Sensing Suite}
The angular acceleration disturbance we can expect from sloshing is given by $I_{sat}\dot{\Omega}_s = \Gamma_s$.
So, by integration, the CFD results show that the typical angular velocity disturbance we can expect from sloshing to be $0.073 \degree/s < \Omega_s < 0.388 \degree/s$. Table \ref{table:max_press_Omega} further summarizes the detection thresholds for detecting pressure changes ($\Delta P$) and angular velocity changes ($\Omega_s$) caused by the slosh disturbance.

\begin{table}[H]
\caption{\label{table:max_press_Omega} Pressure Sensor and Angular Velocity Detection Thresholds }
\centering
\begin{tabular}{cccc}
\hline
$\Gamma_{RW}$ Direction & Max $\Delta$ P(psi) &  Max $\dot\Omega_s$ (\degree $/s^2$) & Max $\Omega_s$ (\degree /s) \\ \hline
X-Axis  & 0.276  & 5.93    & 0.296\\
Y-Axis  & 0.054  & 1.47    & 0.073\\
Z-Axis  & 0.086  & 4.32    & 0.216\\
XYZ-Axis& 0.347  & 7.77    & 0.388\\
\hline
\end{tabular}
\end{table}

Given the angular disturbances of $0.073 \degree/s < \Omega_s < 0.388 \degree/s$, we will follow the treatment of Markley \cite{Crassidis} to ensure that a typical gyroscope can measure the anticipated sloshing disturbance with the Blue Canyon Technologies XACT-50 ADCS. Gyroscopic noise consists of two components and are typically provided by the gyroscope manufacturer:

\vspace{-0.5cm}
\begin{equation}
    \sigma_v = \text{Angular Random Walk:} \text{units of } \degree / \sqrt{\text{time}}
    \label{ang_random_walk}
\end{equation}

\vspace{-1.00cm}
\begin{equation}
    \sigma_u = \text{Gyro Bias Instability:} \text{units of } \degree / time^{3/2}
    \label{bias_stability}
\end{equation}
\vspace{-1.00cm}

We seek to determine the measurement noise (standard deviation), $\sigma_{\Omega}$, for the gyroscopes in the BCT XACT-50 ADCS and compare this value to the previously determined sloshing angular disturbance, $\Omega_s$ over a measurement interval, $\Delta t$.

\begin{equation}
    \sigma_{\Omega} = \frac{\sigma_v^2}{\Delta t} + \frac{1}{3}\sigma_u^2\Delta t
    \label{sigma_std_dev}
\end{equation}

At the time of this writing, BCT sensitivity numbers were not available, so we estimate gyro noise based on the Epson M-G364PDCA sampled at 100Hz. These numbers will be revised upon receiving more details from Blue Canyon Technologies.

\begin{equation}
    \sigma_v = 0.09 \degree/\sqrt{\text{hr}} = 0.0015 \degree/\sqrt{\text{s}}.)
    \label{sigma_v}
\end{equation}

\begin{equation}
    \sigma_u = \frac{\num{6.11e04}\degree/s}{\sqrt{500 s}} \degree/s^{3/2} = \num{2.7e-5} \degree/s^{3/2}
    \label{sigma_u}
\end{equation}

Plugging in these values into Eq. \ref{sigma_std_dev}:

\begin{equation}
    \sigma_{\Omega}^2 = \frac{0.0015\degree/\sqrt{s}}{0.01 s} + \frac{1}{3}(\num{2.7e-5}\degree/s^{3/2})(0.01 s) = \num{2.25e-4}\degree/s^2
\end{equation}

Therefore $\sigma_{\Omega} = \sqrt{\num{2.25e-4}} = 0.015 \degree/s$. Assuming the BCT XACT-50 compares well to these Epson M-G364PDCA noise characteristics, this $\sigma_{\Omega} = 0.015 \degree/s$ detection threshold exceeds the sloshing angular disturbances we model in Table \ref{table:max_press_Omega} which ranges from $0.073 - 0.388 \degree/s$. BCT data sheets report a pointing accuracy of 10.8 to 25.2 arcseconds (0.003 - 0.007 deg).

\begin{figure}[H]
    \centering
    \includegraphics[width=14cm]{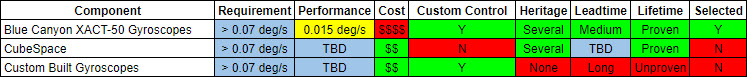}
    \caption{\modif{SPICEsat ADCS Trade Study}}
    \label{fig:ADCS_Trade}
\end{figure}

\modif{Although this study indicates there is less than a factor of ten in the detection threshold of the sensors compared to the disturbance, past analysis by the SloshSat FLEVO experiment in 2005 which used only MSS data indicated that the effect is measurable. SPICEsat will also be able to correlate the MSS data and measured disturbances with the visual data from the camera.  
\\
A critical requirement for SPICEsat's ADCS is to allow SPICEsat to control the internal reaction wheels. The trade study revealed that of the two suitable commercially available ADCS systems that fit the mission, only BCT or a custom-built ADCS would be suitable. See Fig. \ref{fig:ADCS_Trade} for the applicable trade study on ADCS selection.}

\subsubsection{\modif{Liquid Sensing Suite}}

\begin{figure}[H]
    \centering
    \includegraphics[width=14cm]{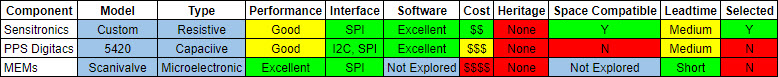}
    \caption{\modif{SPICEsat Pressure Sensor Trade Study}}
    \label{fig:Pressure_Trade}
\end{figure}

The CFD results also show that the typical pressure disturbance we can expect from sloshing is 0.054 psi < $\Delta P$ < 0.347 psi. See Table \ref{table:max_press_Omega}. \modif{Three types of pressure sensors were examined for SPICEsat: capacitive, resistive, and microelectronic mechanical systems (MEMS). Initial trade studies indicated that commercial off-the-shelf capacitive sensors were best suited for our application. However, after commencing studies with capacitive pressure sensors, it was quickly determined such sensors require the two capacitive layers to compress and expand based on the amount of force placed on the upper layer, requiring the expulsion of the material between them (venting). These would not be suitable SPICEsat for this reason. See Fig. \ref{fig:Pressure_Trade} for the applicable trade study on pressure sensor selection.}

\modif{SPICEsat selected the Sensitronics pressure sensors for the spacecraft and has begun testing with these. Sensitronics sensors are expected to have an activation threshold in the 0.01-0.02 psi range, which is sufficient to measure the disturbance} 

\subsection{Vision Sensing Suite}
\begin{figure}[H]
    \centering
    \includegraphics[width=14cm]{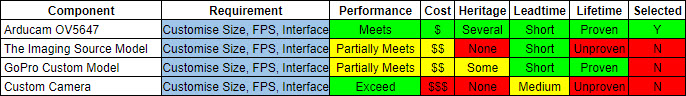}
    \caption{\modif{SPICEsat Camera Trade Study}}
    \label{fig:Camera_Trade}
\end{figure}

The results presented in Fig. \ref{fig:CFD_Results} validate the theory that the vision sensing suite (VSS) on-board SPICEsat will record significant fluid movement on video. CFD animations of the fluid sloshing dynamics clearly show robust fluid movement when $\Gamma_{RW}$ is imparted around the x, z, and xyz axes. Strong fluid movement will be critical in the post-actuation analysis of the fluid using computer vision techniques. The CFD work demonstrates the potential usefulness of the video to be recorded. \modif{Various custom and commercial off-the-shelf cameras were reviewed that met the requirements outlined in Table \ref{table:video_params}. Fig. \ref{fig:Camera_Trade} shows the applicable trade study for camera selection.}

\section{Conclusions}
SPICEsat promises to be a very complicated and interesting mission with the capability to validate existing models and provide a new method for attitude control in space. Previous experiments in this area lack a detailed understanding of internal tank dynamics, which we propose to detect using internal pressure sensors and camera footage. This data will be used to validate equivalent mechanical models and create new, more representative ones. With the new insight, new control algorithms will be trained using classical methods or machine learning before being updated on board. 

In the application of astronomical imaging and military observatory satellites with large propellant tanks, this research translates to faster slew rates, more agile maneuvers, and ultimately, higher performance. Development and implementation of new control methods using Machine Learning to solve the problem of fuel sloshing will directly benefit future aerospace platforms subjected to similar kinds of disturbances and modeling uncertainties.

We have demonstrated that the design of SPICEsat and the payload measurements using the MSS, VSS, and LSS are well within detection limits, placing upper and lower bounds on the sensitivity of each sensing suite.

Future work includes the construction of a "flat sat" in which each payload component will be connected, tested, and then measurements made to ensure proper satellite design. The software and hardware will be tested to ensure compatibility and suitability against the mission requirements. 

\vspace{1cm}
\textbf{Acknowledgements:} The authors wish to thank the United States Air Force Research Lab University Nanosatellite Program for funding this research through award number CP0072457. The authors also wish to thank Dr. Robert Manning of NuSpace for valuable counseling on computational fluid dynamics modeling.

\bibliography{references}

\end{document}